\documentclass[twocolumn]{aastex631}
\usepackage{booktabs}
\usepackage{graphicx}
\usepackage{multirow}
\usepackage{hyperref}

\newcommand\feh{{\rm [Fe/H]}}
\newcommand\logg{{\rm log}\,g}
\newcommand\teff{T_{\rm eff}}
\newcommand\ebv{E(B-V)}
\newcommand\gaia{Gaia}
\newcommand\RV{R_{\rm V}}
\newcommand\AV{A_{\rm V}}
\newcommand\degree{^{\circ}}

\shorttitle{$\RV$ map of the Milky Way}
\shortauthors{Zhang, Yuan \& Chen}
\graphicspath{{./}{figures/}}

\begin{document}

\title{A $\RV$ map of the Milky Way revealed by LAMOST}

\author[0000-0003-1863-1268]{Ruoyi Zhang}
\author[0000-0003-2471-2363]{Haibo Yuan}
\affil{Institute for Frontiers in Astronomy and Astrophysics, Beijing Normal University, Beijing 102206, China;  yuanhb@bnu.edu.cn}
\affil{Department of Astronomy, Beijing Normal University, No.19, Xinjiekouwai St, Haidian District, Beijing, 100875, China}

\author[0000-0003-2472-4903]{Bingqiu Chen}
\affil{South-Western Institute for Astronomy Research, Yunnan University, Kunming 650500, China; bchen@ynu.edu.cn} 

\begin{abstract}
The total-to-selective extinction ratio, $\RV$, is a key parameter for tracing the properties of interstellar dust, as it directly determines the variation of the extinction curve with wavelength. By utilizing accurate color excess measurements from the optical to the mid-infrared range, we have derived $\RV$ values for approximately 3 million stars from the LAMOST data release 7 (DR7) using a forward modeling technique. This extensive dataset enables us to construct a comprehensive two-dimensional $\RV$ map of the Milky Way within the LAMOST footprint at a spatial resolution of $\sim$ 27.5\,arcmin. 
Based on reliable sightlines of $\ebv > 0.1$, we find that $\RV$ exhibits a Gaussian distribution centered around 3.25 with a standard deviation of 0.25. 
The spatial variability of $\RV$ in the Galactic disk exhibits a wide range, spanning from small scales within individual molecular clouds to large scales up to kiloparsecs. A striking correlation is observed between the distribution of $\RV$ and molecular clouds. Notably, we observe lower $\RV$ values within the regions of nearby molecular clouds compared to their surrounding areas.
Furthermore, we have investigated the relationships between $\RV$ and various parameters, including dust temperature, dust emissivity spectral index, column density of atomic and molecular hydrogen, as well as their ratios and the gas-to-dust ratio. We find that these relationships vary with the level of extinction. These analyses provide new insights into the properties and evolution of dust grains in diverse interstellar environments and also hold significant importance for achieving accurate extinction corrections.
\end{abstract}

\keywords{ISM: dust, extinction --- stars: general --- molecular clouds.}

\section{Introduction} \label{sec:intro}

The extinction law, also known as the extinction curve, represents the variation of extinction with wavelength or frequency.  
It plays a crucial role in correcting the effects of dust extinction in observed objects and has been extensively studied to understand the properties of dust grains.
\citet{1986ApJ...307..286F,1988ApJ...328..734F} parameterized the extinction curve in the ultraviolet (UV) bands using a simple 6-parameter function.
\citet{1989ApJ...345..245C} showed that the extinction curve from the UV to the optical (303\,nm -- 3.5\,$\mu$m) can be described by a one-parameter family of curves with different values of the parameter $\RV$. 
The $\RV$ parameter, which is known as the total-to-selective extinction ratio, is defined as $\RV \equiv \AV / \ebv$.
Much (but not all) of the spatial variability seen in extinction curves has been shown to correlate with $\RV$.
Consequently, this one-parameter description allows us to predict the extinction value at any wavelength within the optical to UV range, based solely on the knowledge of the $\RV$ value along a particular line of sight.
A smaller $\RV$ value indicates a steeper extinction curve, implying a larger difference in extinction at different wavelengths.
Since then, numerous studies have been conducted to model extinction curves, employing various fitting techniques and considering different sightlines and wavelength ranges \citep[e.g.,][]{1990ApJS...72..163F, 1994ApJ...422..158O, 2014A&A...564A..63M, 2019ApJ...886..108F, 2023arXiv230401991G}.

$\RV$ not only serves as a parameter to describe extinction curves but also provides valuable information about the properties of dust grains.
The model proposed by \citet{2001ApJ...548..296W} demonstrated a correlation between larger average grain sizes and higher $\RV$ values.
However, other factors such as the chemical composition of dust grains may also contribute to the variation in $\RV$ values. Observations of the diffuse interstellar medium in the Milky Way suggest an average $\RV$ value of approximately 3.1 \citep{1979ARA&A..17...73S,1989ApJ...345..245C}, but the variation in different sightlines can be very large.
It is widely accepted that sightlines passing through dense molecular clouds with high extinction tend to exhibit higher $\RV$ values, reaching values as large as approximately 6\citep[e.g.,][]{1999PASP..111...63F}.
This variation can be attributed to dust grain growth through processes such as accretion and coagulation \citep[e.g.,][]{1984ApJ...283..123V, 1999PASP..111...63F, 2003ARA&A..41..241D, 2012A&A...548A..61K, 2013MNRAS.428.1606F}.
In low-density regions, $\RV$ values could be as small as approximately 2 \citep[e.g.,][]{1992ApJ...393..193W,1999PASP..111...63F, 2017ApJ...848..106W}. 
When dust grains are fully exposed to radiation, they become more susceptible to destruction through processes such as sputtering by impinging atoms or ions, photolysis by UV photons, and photodesorption occurring on the dust surface \citep{2003ARA&A..41..241D}.

Prior to the 2010s, the number of O and B stars with accurate $\RV$ measurements was limited to a few hundred \citep[e.g.][]{2007ApJ...663..320F}.
However, the availability of precise stellar parameter measurements and intrinsic colors from modern large-scale astronomical surveys has opened up the possibility of measuring $\RV$ values for large samples of stars across extensive sky regions.
\citet[hereafter S16]{2016ApJ...821...78S} measured $\RV$ values along 150,000 sightlines based on the APOGEE spectroscopic data and a series of photometric bands from the optical to the near-infrared (near-IR).
They mapped the distribution of $\RV$ in the nearby Galactic mid-plane (within a distance of $<$ 4 kpc) and found that most variations in $\RV$ are not correlated with dust column density in the range of $0.5 < \ebv < 2$, but instead show a correlation with distance.
Furthermore, S16 discovered a strong negative correlation between $\RV$ and the dust emissivity spectral index $\beta$ between 353 and 3000\,GHz, which suggests that conditions that lead to steep far-infrared emission spectra also lead to optical and infrared extinction curves become steeper.
However, due to the sparse sampling of diverse environments, the detailed physical and chemical mechanisms underlying the variations in $\RV$ are still unclear.

The Large Sky Area Multi-Object Fiber Spectroscopy Telescope (LAMOST; \citealt{2012RAA....12..723Z, 2012RAA....12.1197C,2014IAUS..298..310L}) has provided us with over 10 million high-quality stellar spectra and precise atmospheric parameters, enabling us to examine the spatial variability of $\RV$ with greater spatial resolution and wider sky coverage. In our previous study \citep[hereafter ZY23]{2023ApJS..264...14Z}, we accurately measured the reddening for 5 million stars across a wide range of wavelengths, achieving typical errors of 0.01-0.03\,mag for individual colors. 

Based on the results, we aim to construct a highly precise $\RV$ map that spans from the Galactic disk to the halo by utilizing the extinction law of \citet[hereafter F99]{1999PASP..111...63F} and the BOSZ synthetic spectral database \citep{2017AJ....153..234B}. This map will enable us to investigate the physical properties of dust in unprecedented detail and provide valuable observational constraints on the formation and evolution of dust grains. 

The paper is organized as follows. 
In Section \ref{sec:Data}, we briefly introduce the adopted data sets. 
In Section \ref{sec:Method}, we obtain $\RV$ of each star by a forward modeling approach and check the reliability of the results. 
In Section \ref{sec:Rv_map}, we present our $\RV$ map and discuss its implications. 
In Section \ref{sec:Discussion}, we compare the measurements with the study in the literature, investigate the spatial coincidence between $\RV$ and molecular clouds, and study the correlation between $\RV$ and some parameters of the interstellar medium (ISM). 
We summarize our findings in Section \ref{sec:Summary}.

\section{Data} \label{sec:Data}

The data used in this study were obtained from our previous work, ZY23, which provided highly accurate reddening measurements for approximately 5 million stars across 21 different colors. The typical errors of the reddening values are only 0.01 -- 0.03\,mag, depending on color. For this study, we have selected specific colors from various passbands spanning the optical to mid-infrared wavelength range. These passbands include the $g/r/i/z/y$-band of Pan-STARRS 1 (PS1; \citet{2018AAS...23110201C}), the $G_{\rm BP}/G/G_{\rm RP}$-band of $\gaia$, the $u/g/r/i/z$-band of Sloan Digital Sky Survey (SDSS; \citet{2015ApJS..219...12A}), the $J/H/K_{\rm S}$-band of Two Micron All Sky Survey (2MASS; \citet{2006AJ....131.1163S}), and the $W1/W2$-band of Wide-field Infrared Survey Explorer (WISE; \citet{2010AJ....140.1868W}). We excluded the passbands from the Galaxy Evolution Explorer (GALEX; \citet{2005ApJ...619L...1M}) and the $W3/W4$-band of WISE due to the limited number of sources with reliable photometric quality. 

The stellar parameters used in the paper, including effective temperatures $\teff$, surface gravities $\logg$, and metallicities $\feh$, have been derived from the LAMOST Stellar Parameter Pipeline (LASP) \citep{2011RAA....11..924W, 2015RAA....15.1095L} and the HotPayne catalog of \citet{2022A&A...662A..66X}.
The $\ebv$ values for each star are calculated using their corresponding $E(G_{\rm BP}-G_{\rm RP})$ and $R(G_{\rm BP}-G_{\rm RP})$ values, which were obtained from ZY23. Since the values of $R(G_{\rm BP}-G_{\rm RP})$ used in this study were obtained using the $\ebv$ values from the dust map of \citet[hereafter SFD]{1998ApJ...500..525S} in the intermediate to high Galactic latitude regions, we have corrected for the 14\% overestimation of SFD $\ebv$ values \citep{2010ApJ...725.1175S,2013MNRAS.430.2188Y}.

Firstly, we utilize the same sample cleaning and trimming criteria as described in the data section of ZY23.
Then, we binned the stars in our dataset by their values of $\ebv$ and $\teff$.
Subsequently, we performed a 3$\sigma$ rejection of the reddening values for each color within each bin in the corresponding color excess - $\ebv$ diagram.
These mitigate the impact of inaccurate reddening measurements in the subsequent analysis.
On average, this step removes approximately 10--20\% of color excesses.

\section{Method} \label{sec:Method}

To create the $\RV$ map, we must initially determine the $\RV$ value for each star in our sample. To achieve this, we used a forward modeling technique that compares simulated color excess ratios (CERs) with measured values to determine each star's best-fit $\RV$. 

After verification, we find that differences in different dust models can lead to changes in the overall distribution of the sample's measured $\RV$, but have little impact on the relative difference of $\RV$. We compared the measured data with F99 and the models of \citet{2019ApJ...886..108F} and \citet{2023ApJ...950...86G}. Our preliminary analysis shows that the F99 model fits better for the color excess of this sample, and therefore the F99 model is used for calculations in the follow-up process. It is worth noting that the evaluation of other extinction laws requires a dedicated study, but this is beyond the scope of this paper.

We adopted the F99 reddening law to characterize the $\RV$-dependent extinction curves, expressed as $A_\lambda/\AV$, where $A_\lambda$ represents the extinction at a specific wavelength $\lambda$, and $\AV$ is the V-band extinction. The F99 reddening law has been widely used in literature and is consistent with observational data, as demonstrated in previous studies such as \citet{2011ApJ...737..103S} and \citet{2013MNRAS.430.2188Y}.
For a given star, the extinction at a specific passband $x$ can be expressed as,
\begin{equation}  \label{equ1}
    A_x = -2.5 \times \log 
    \left (\frac{\int F_0(\lambda) \cdot S(\lambda) \cdot R(\lambda) \cdot \lambda / hc\ d\lambda} 
    {\int F_0(\lambda) \cdot S(\lambda) \cdot \lambda / hc\ d\lambda} \right ),
\end{equation}
where $\lambda$ is the wavelength, $h$ is the Planck constant, $c$ is the speed of light, $F_0(\lambda)$ is the intrinsic flux of the star, $S(\lambda)$ the filter response curve of the passband $x$, and $R(\lambda)$ the extinction term. According to the F99 extinction law, $R(\lambda)$ can be expressed as,
\begin{equation}  \label{equ2}
    R(\lambda) = 10 ^{ -0.4 \times A_\lambda} = 10 ^{ -0.4 \cdot \frac{A_\lambda}{\AV} \cdot R_{\rm V} \cdot E(B-V)}.
\end{equation}
In this study, we utilized the intrinsic flux $F_0(\lambda)$ from the BOSZ stellar atmosphere models database \citep{2017AJ....153..234B} and the filter profiles $S(\lambda)$ of the individual passbands from \citet{2020sea..confE.182R}. 
We only used the BOSZ spectra with stellar parameters of $\feh$ = 0, [$\alpha$/H] = 0, and [C/H] = 0, 
as varying $\feh$, [$\alpha$/H], or [C/H] has negligible impact on the results. 
Additionally, to obtain templates of $A_x$, we adopted a grid that covers the following parameter ranges: $\teff$ from 3750 to 20000\,K in steps of 250\,K for $\teff \le$ 7500\,K and 500\,K for $\teff \ge$ 7500\,K, $\ebv$ from 0 to 3\,mag, with a step size of 0.01\,mag for $\ebv < 0.2$\,mag, 0.04 mag for $0.2 < \ebv < 1$\,mag, and 0.25 mag for $\ebv > 1$\,mag, $\logg$ from 0 to 5 with a step size of 0.5, and $\RV$ from 1 to 7 with a step size of 0.1.

\begin{figure*}
    \centering
    \includegraphics[width=\linewidth]{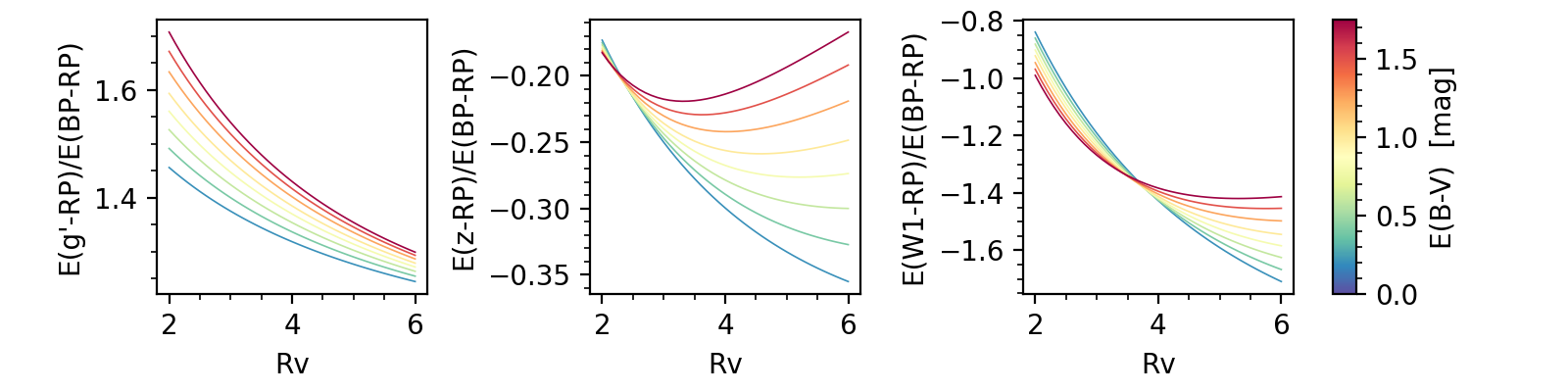}
    \includegraphics[width=\linewidth]{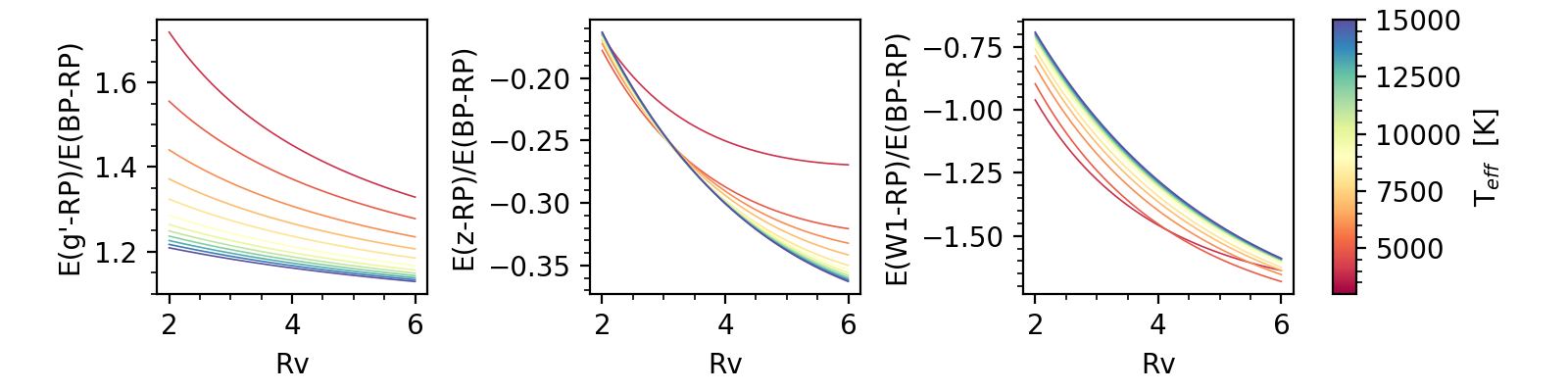}
    \caption{{\it Top panels}: the simulated CERs plotted as a function of $\RV$ and $\ebv$, with a fixed $\teff$ of 5500\,K and $\logg$ of 4. {\it Bottom panels}: the simulated CERs as a function of $\RV$ and $\teff$, with a fixed $\ebv$ of 0.4\,mag and $\logg$ of 4.
    }
    \label{fig:modelCER}
\end{figure*}

To derive the best-fit $\RV$ values, we simulated the CERs using the expression $E(x-RP)/E(BP-RP)= ({A_x- A_{RP})/(A_{BP}-A_{RP}})$, and then compared them with observations. This form of CER was used because we lack absolute extinction measurements for individual stars and only have reddening data. 
Among the photometric bands in our dataset, $\gaia$ photometry provides the largest sample size and the highest degree of accuracy. 
In addition, our previous research has shown that the model reddening coefficients for the $\gaia$ filters align well with empirical coefficients, thereby reducing potential uncertainties.
In Fig.~\ref{fig:modelCER} we show the simulated CERs as functions of the $\RV$, $\ebv$, and $\teff$ parameters. We note that CERs for $z$ and $W1$ filters at $\teff$ = 4000\,K (the reddest lines in the bottom middle and right panels, respectively) appear abnormal, probably due to the potential issues with the BOSZ model in those IR bands and at low temperature.

\begin{figure*}
    \centering
    \includegraphics[width=\linewidth]{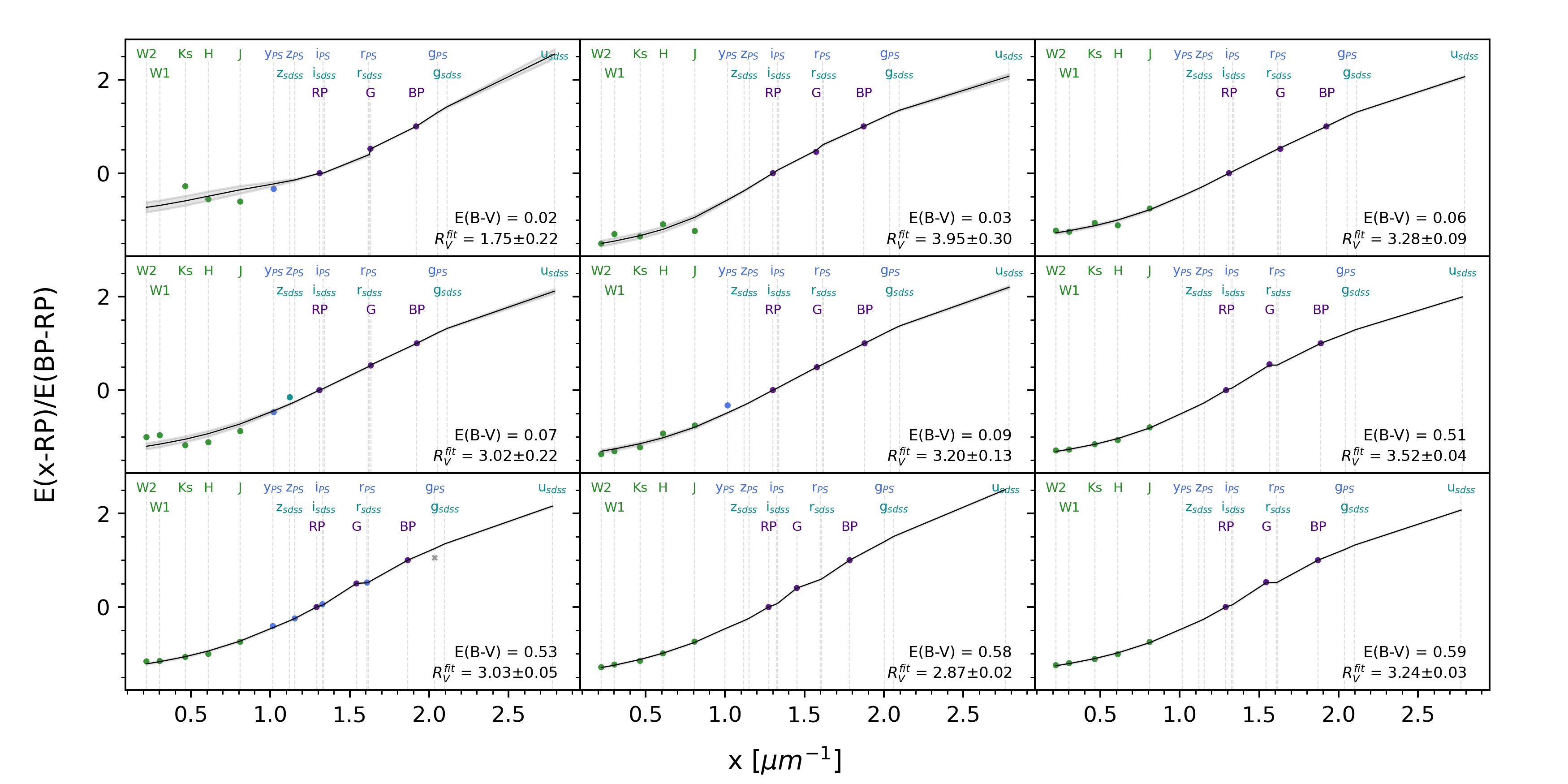}
    \caption{Nine examples of the fitting of $\RV$. The gray shading region of each passband represents the error range of the simulated CER. }
    \label{fig:example}
\end{figure*}

We used the linear least squares estimation method {\sc curve\_fit} of the {\sc SciPy} module \citep{2020SciPy-NMeth} to obtain the best-fit $\RV$ values for individual stars in our sample.
We first identified the nearest grid point to the measured $\teff$ and $\ebv$. Subsequently, we performed a linear interpolation of the $\RV$ grid points to find the optimal $\RV$ value that minimized the differences between the measured and simulated CERs. During this process, we also estimated the standard deviation error, denoted as $err(\RV)$. In our analysis, only stars with a minimum of five CERs (including both the $\gaia$ $BP$ and $RP$ bands) were considered.  Fig.\,\ref{fig:example} illustrates several examples of the $\RV$ fits. Overall, the model CERs exhibit a good agreement with the measured values. Furthermore, the dispersion and estimated errors decrease with increasing extinction.

We have constructed mock data to test our method.
The simulation revealed there is a systematic underestimation of the $\RV$ values obtained at low extinction. 
For example, for mock samples with $\RV$ of 3.1 and corresponding $\ebv$ values of 0.05 and 0.02, the measured $\RV$ would be underestimated by 0.2 and 1.4, respectively.
Subsequent analysis indicated that such systematic errors become negligible when $\ebv \ge 0.1$, and we defined this threshold as the criterion for the `reliable' sample.
Consequently, it is essential to develop a more refined approach to obtain well-measured $\RV$ values at low extinction regions in the future.

\begin{figure}
    \centering
    \includegraphics[width=\linewidth]{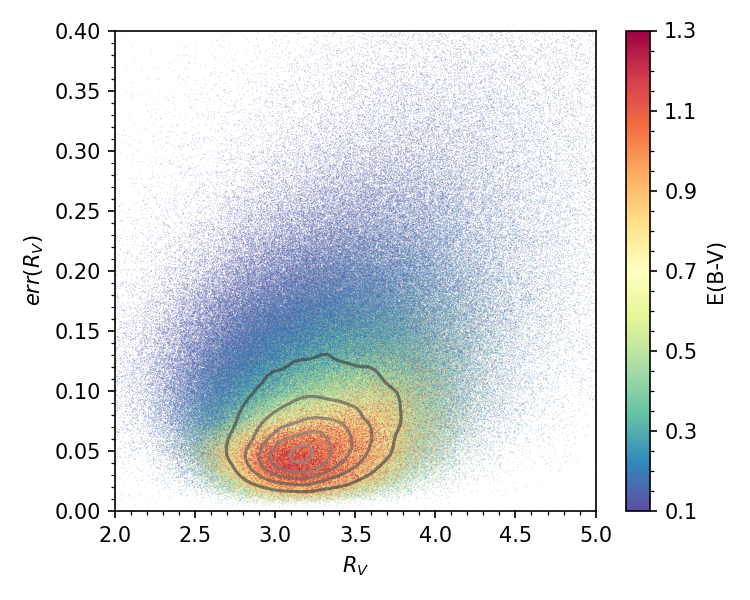}
    \caption{The $\RV$-$err(\RV)$ distribution of the reliable sample. Contour lines represent the density of the cataloged stars on a linear scale, and the color of each point corresponds to its $\ebv$ value.}
    \label{fig:Rv&err}
\end{figure}

Finally, we have obtained a total of 3,182,038 valid $\RV$ measurements, out of which 1,178,719 are considered reliable.
The values and errors of the reliable sample are displayed in Fig.~\ref{fig:Rv&err}. 
The density contours reveal a concentration of results around $\RV=3.2$ and $err(\RV)=0.05$. 
Notably, there is a positive correlation between $\RV$ and $err(\RV)$, which is more pronounced for stars with high $\ebv$. 
This correlation arises primarily from the increased sensitivity of the F99 extinction curve to changes in $\RV$ as $\RV$ values increase, consequently causing larger $err(\RV)$.

\begin{figure*}
    \centering
    \includegraphics[width=\linewidth]{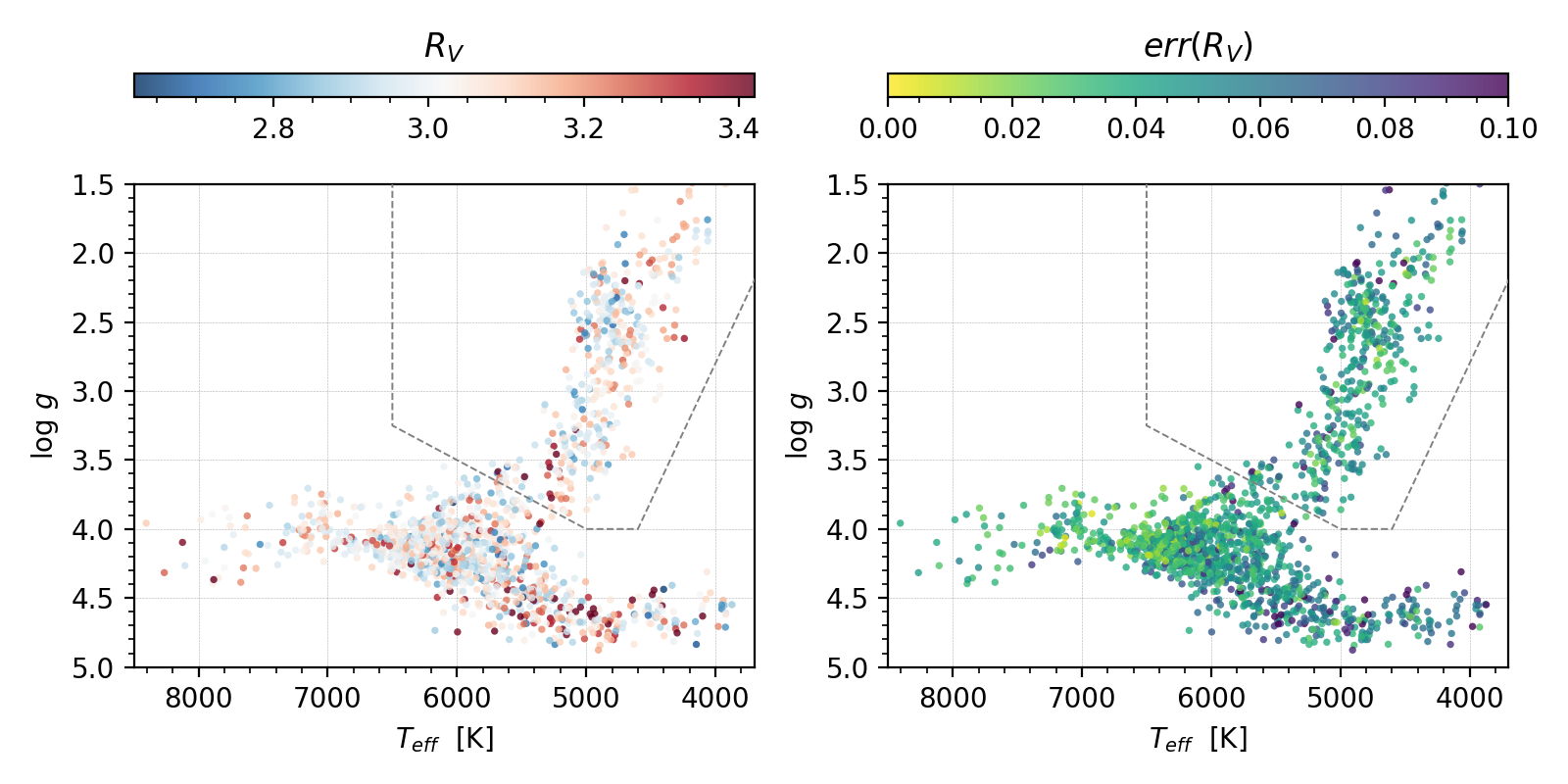}
    \caption{$\RV$ (left panel) and $err(\RV)$ (right panel) distributions in the $\teff$ versus $\logg$ diagram for a small sample of stars selected based on their longitude ($172.5\degree < l < 174.5\degree$), latitude ($-22\degree < b < -20.5\degree$), and $\RV$ error ($err(\RV) < 0.2$). The dashed lines mark the boundary dividing the giants from the dwarfs.}
    \label{fig:HRD}
\end{figure*}

\begin{figure*}
    \centering
    \includegraphics[width=\linewidth]{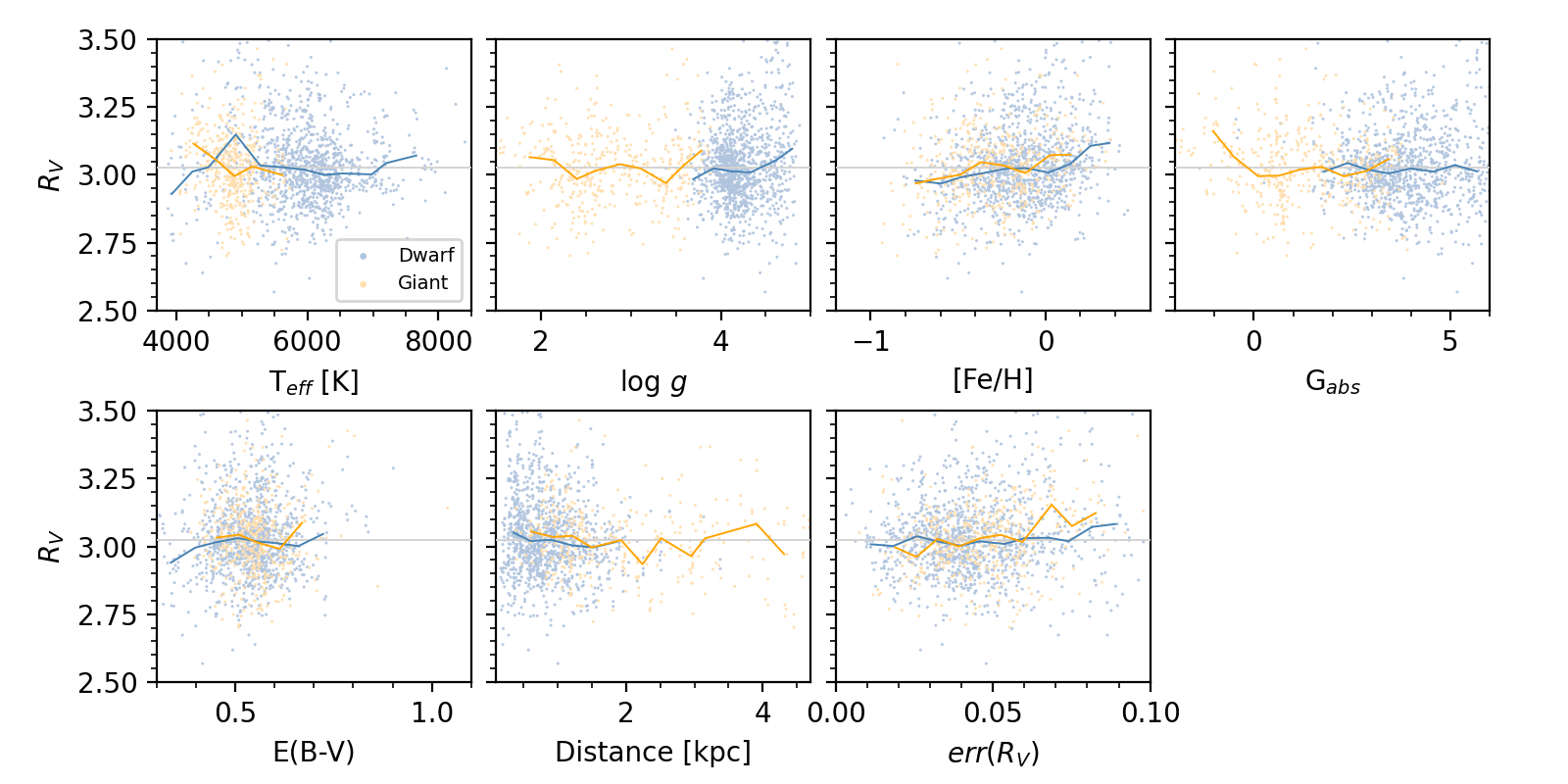}
    \caption{The correlation between $\RV$ values and stellar parameters for stars in Fig.~\ref{fig:HRD}. Dwarf and giant stars are represented by blue and yellow dots, respectively, and the median values for the binned data are shown by corresponding colored lines. The median $\RV$ is represented by a solid gray line.}
    \label{fig:checkout}
\end{figure*}

To investigate potential correlations between best-fit $\RV$ values and stellar parameters, we have selected stars located in a specific region at the edge of the Taurus complex ($l$ = 172.5 -- 174.5\degr\ and $b$ = $-22$ -- $-20.5$\degr). 
After excluding stars with $err(\RV) > 0.1$, we have plotted the distribution of $\RV$ and $err(\RV)$ in the $\teff$-$\logg$ diagram, as shown in Fig.~\ref{fig:HRD}.
Furthermore, we have examined the correlations between $\RV$ and several stellar parameters, including $\teff$, $\logg$, [Fe/H], $M_G$, $\ebv$, $d$, and $err(\RV)$, which are displayed in Fig.~\ref{fig:checkout}.  
$M_G$ values are calculated using the $G$-band extinction coefficient, observed $G$ magnitude, and distance from $\gaia$ DR3 \citep{2021A&A...649A...1G}. 
Overall, our findings indicate that $\RV$ values are generally independent of these parameters, as expected. 
However, the $\RV$ values are slightly larger for giant stars at $\teff$ = 4000\,K.
This may be partly attributed to the large uncertainties resulting from the inability of the BOSZ stellar models to accurately predict the near-IR flux of low-temperature stars.

\section{The 2D $\RV$ Map} \label{sec:Rv_map}

\begin{figure*}
    \centering
    \includegraphics[width=\linewidth]{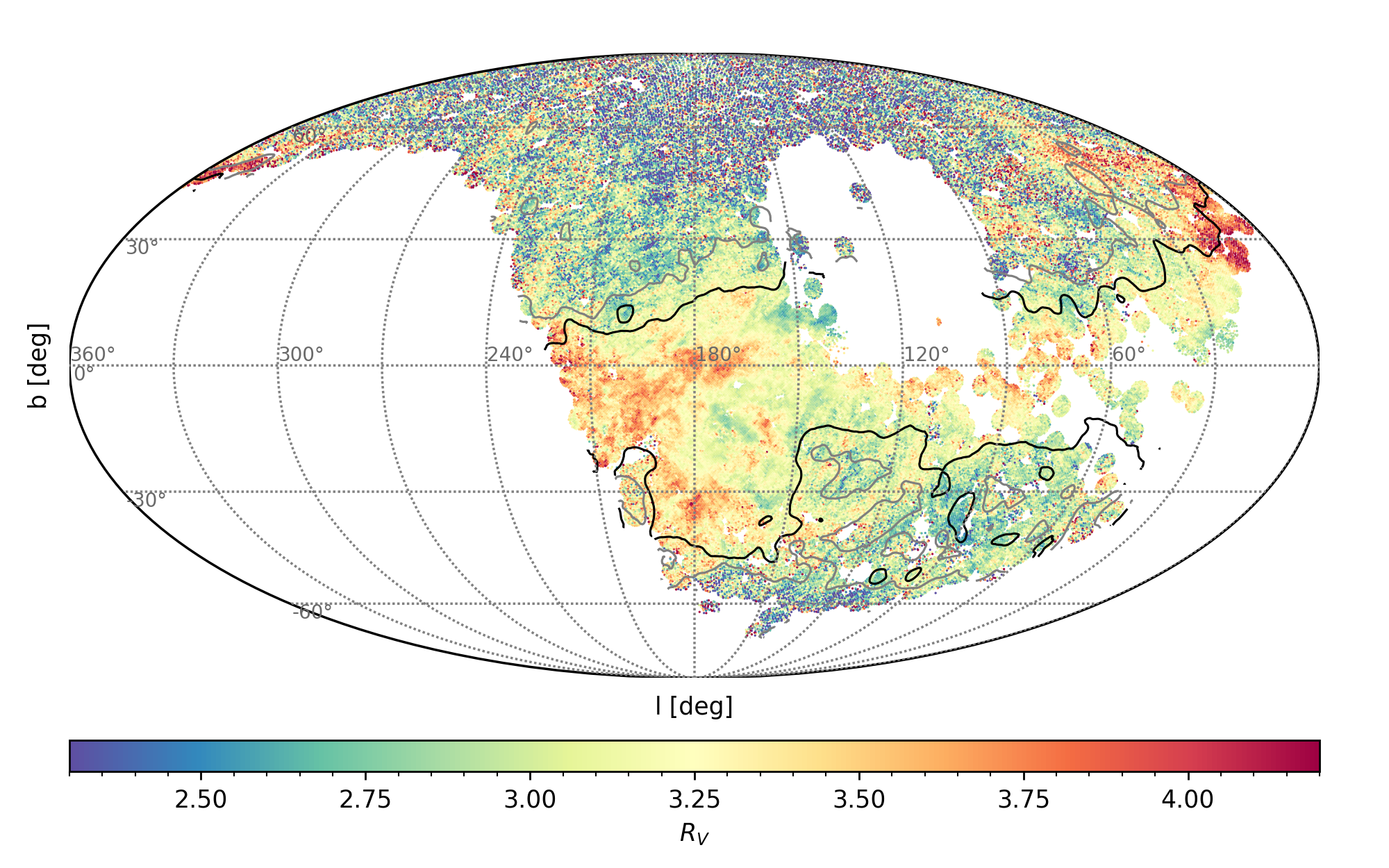}
    \caption{The 2D $\RV$ map. The colored HEALPix points depict the median values of $\RV$ in each direction. Uncolored regions indicate directions with insufficient stars for analysis. The black and gray lines on the map indicate the sightline directions with an extinction value of 0.1 and 0.05, respectively.}
    \label{fig:allsky}
\end{figure*}

To create a 2D map of $\RV$, we divided the validly measured stars into small pixels. In this study, we partition the celestial sphere into 196,608 sightlines by using the HEALPix scheme ($nside = 128$, a spatial resolution of $\sim$ 27.5\,arcmin). After applying a 3-$\sigma$ clipping for each sightline, we calculate the median $\RV$ and $\ebv$ values. We do not use the weighted average algorithm due to the positive correlation between $\RV$ and $err(\RV)$, which can cause systematic underestimations of $\RV$. We also excluded sightlines with less than five sources, which are primarily distributed at high Galactic latitudes. The resulting 2D $\RV$ map is shown in Fig.~\ref{fig:allsky}. 

Fig.~\ref{fig:allsky} presents the distribution of $\RV$ across a wide expanse of our Galaxy, encompassing regions ranging from the Galactic disk to the halo. The $\RV$ measurements in the Milky Way exhibit significant variations across different regions, with values spanning from 2.0 to 4.5. Notably, these values do not follow a random pattern but exhibit small-scale and large-scale variations throughout the Milky Way. Specifically, $\RV$ values are higher in certain regions, such as those with Galactic coordinates of $l = 180 - 210$\degr\ and $b < 10$\degr, $l = 60 - 120$\degr\ and $|b| < 15$\degr, and near $-30 < l < 30$\degr\ and $b = 30$\degr. Conversely, $\RV$ values are lower in regions such as those with coordinates near $l = 140 - 16$0\degr\ and $|b| < 20$\degr, and of $l = 25-60$\degr\ and $b = -20-10$\degr.  Apart from these large-scale patterns, the $\RV$ values also exhibit fine-scale structures down to the scales inside individual molecular clouds.

As shown in Fig.~\ref{fig:allsky}, the distribution of $\RV$ in the low extinction region with a median $\ebv$ less than 0.1 (indicated by black contours) appears to exhibit a high degree of randomness. This is primarily due to the significant errors in the measured CERs resulting from low reddening values. To obtain a robust $\RV$ distribution, it is necessary to eliminate sightlines with large errors. Additionally, the gray lines on the map highlight the transition region with $0.05 < \ebv < 0.1$, indicating that measurements of $\RV$ within these regions should be used with caution. In subsequent analyses, we constructed a 2D map using reliable samples instead of considering all valid measurements. Furthermore, we employ the median absolute deviation (MAD) and standard deviation (STD) values to identify a set of `reliable' sightlines. A sightline is deemed reliable if $mad/\RV < 10\%$, $std/\RV < 15\%$, and the number of reliable stars $N >5$. As a result, 71\% sightlines were eliminated and remained 24,795 reliable sightlines.

\begin{figure}
    \centering
    \includegraphics[width=\linewidth]{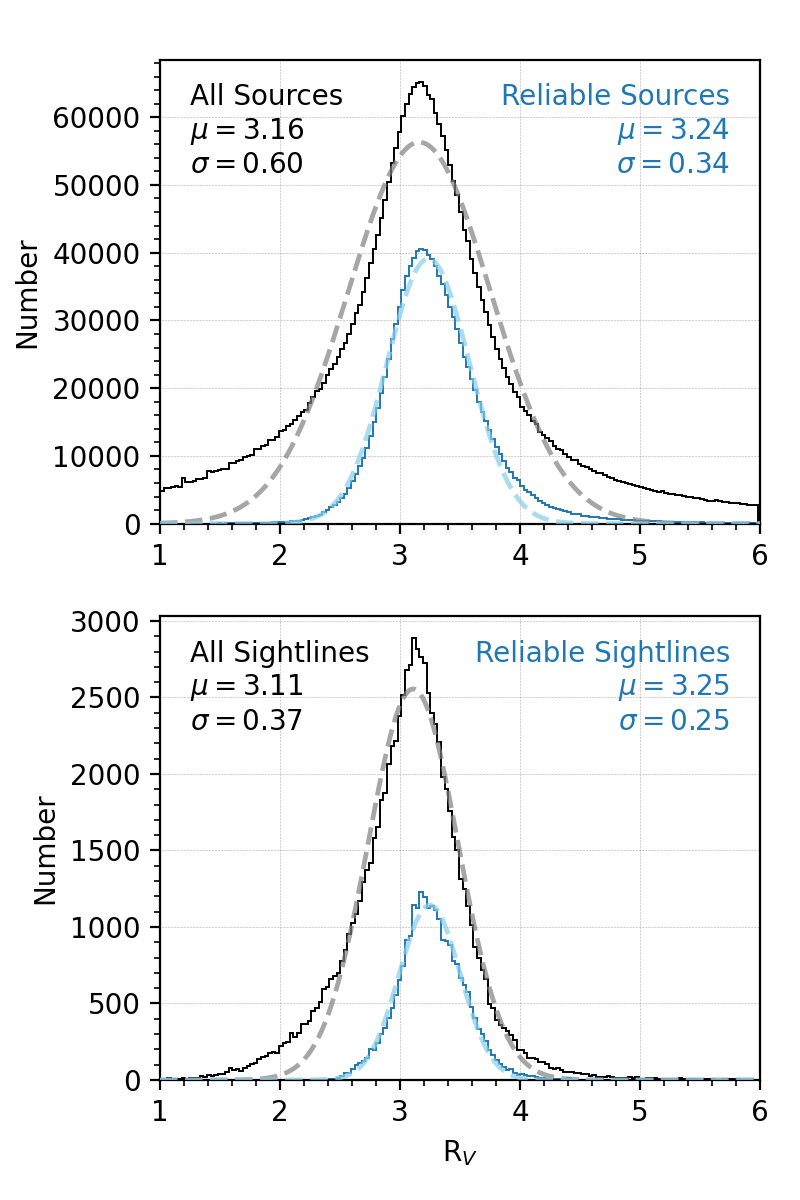}
    \caption{The histograms of the best-fit $\RV$ values for individual stars (upper panel) and the median $\RV$ values for individual sightlines (bottom panel). The black lines in both panels represent the distribution of all sources or all sightlines, while the blue lines represent only reliable sources or sightlines. In each panel, the corresponding dashed line in blue or black represents the best Gaussian fit, and the label indicates the relevant parameters. 
    }
    \label{fig:hist}
\end{figure}

The histograms of the best-fit $\RV$ values for individual stars and the median $\RV$ values for the individual sightlines are shown in Fig.~\ref{fig:hist}. The distribution of $\RV$ for all stars in the catalog is non-Gaussian, which may be partly due to the large measurement error for stars with low extinction, and partly because the intrinsic distribution may well be non-Gaussian.
However, the distribution of $\RV$ values for the reliable sample closely follows a Gaussian shape, with a slight excess in the higher $\RV$ end. The mean $\RV$ for the reliable sample is 3.24, with a dispersion of 0.34. This result slightly differs from the findings of \citet{2016ApJ...821...78S} (who reported $\mu=3.32$ and $\sigma=0.18$) due to variations in the $\RV$ calculation methods and the traced regions, as discussed in Section~\ref{sec:Discussion}. The median $\RV$ values for individual sightlines show a narrower distribution in the bottom panel of Fig.~\ref{fig:hist}.
For the selected reliable sightlines, the distribution is well described by a Gaussian with a mean of 3.25 and a standard deviation of 0.25.

\section{Discussion} \label{sec:Discussion}

\subsection{Comparison with \citet{2016ApJ...821...78S}} \label{sec:Comparison}
\begin{figure*}
    \centering
    \includegraphics[width=\linewidth]{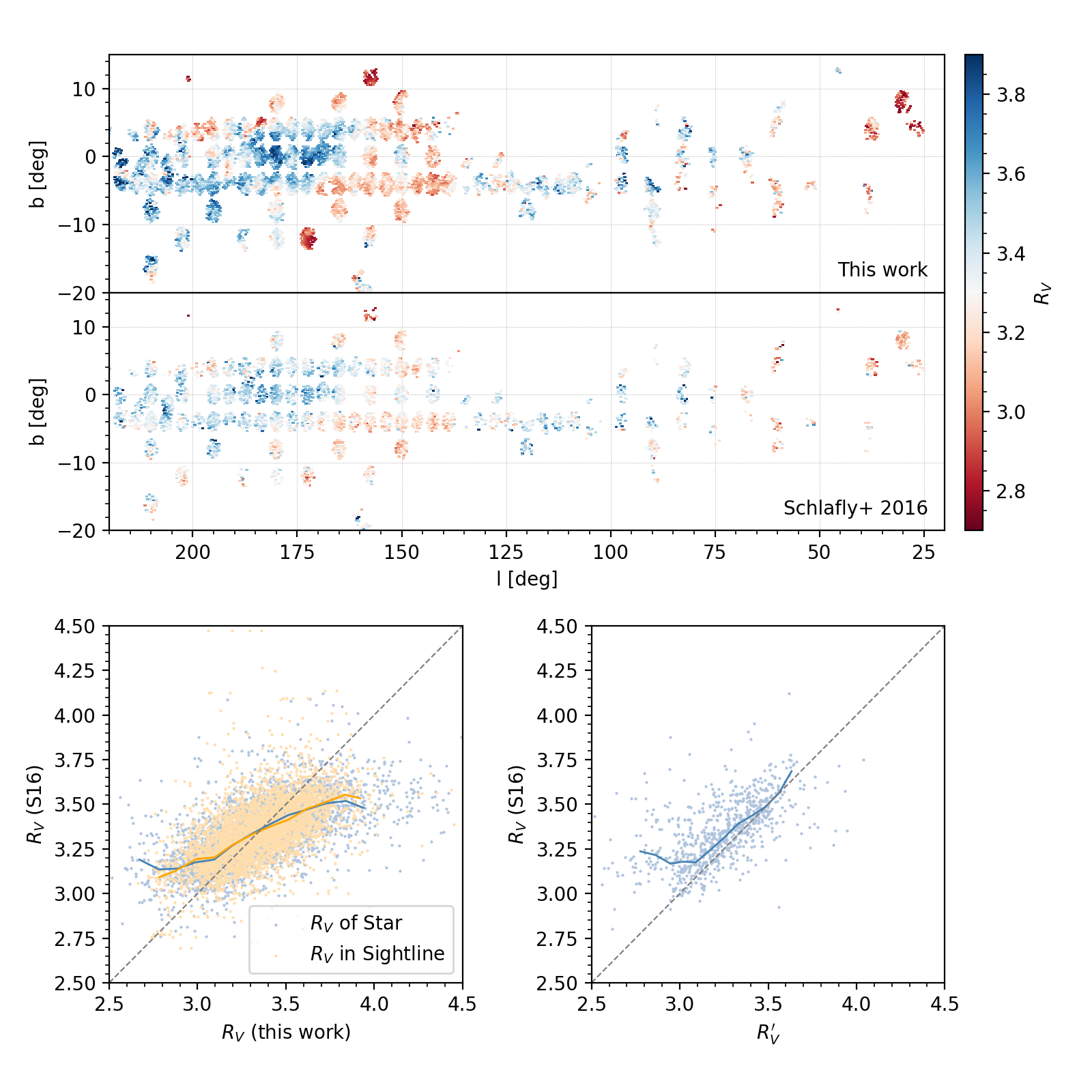}
    \caption{A comparison of $\RV$ values between our work and \citet{2016ApJ...821...78S}. In the upper panel, we show the spatial comparison. The $\RV$ values are represented by colored squares, with each square indicating a common sightline. The color of the square corresponds to the value of $\RV$. In the bottom panels, we show the comparisons of $\RV$ values of the individual stars or the individual sightlines. The blue dots indicate the $\RV$ of stars and the yellow dots represent the $\RV$ of the sightlines. The solid lines of the corresponding colors represent the binned median values. The dashed lines representing $y=x$ are plotted to guide eyes.}
    \label{fig:liter}
\end{figure*}

We compare our $\RV$ measurements with those from \citet{2016ApJ...821...78S}. We selected common sources from both our reliable sample and the catalog of S16. Using the HEALpix algorithm, we determined the individual sightlines from the common sources and calculate the median $\RV$ values for both works. The comparison is shown in Fig.~\ref{fig:liter}. The spatial features are in good agreement (top panel). However, there is a linear systematic difference between our $\RV$ values and those from \citet{2016ApJ...821...78S} (bottom left panel), which mainly resulted from different ways of calculating $\RV$. We used the CERs obtained in our results to recalculate new values based on the formula in \citet{2016ApJ...821...78S}, given by $\RV' = 1.2 \times E(g_{P1} - W2)/E(g_{P1} - r_{P1}) - 1.18$. This approach largely eliminated the systematic error, as shown in the bottom right panel of Fig.~\ref{fig:liter}. Our method for calculating individual $\RV$ values involves CERs estimate from more colors and a detailed forward modeling process, thus being more reliable. 

\subsection{$\RV$ variations in molecular clouds} \label{sec:clouds}

\begin{figure*}
    \centering
    \includegraphics[width=\linewidth]{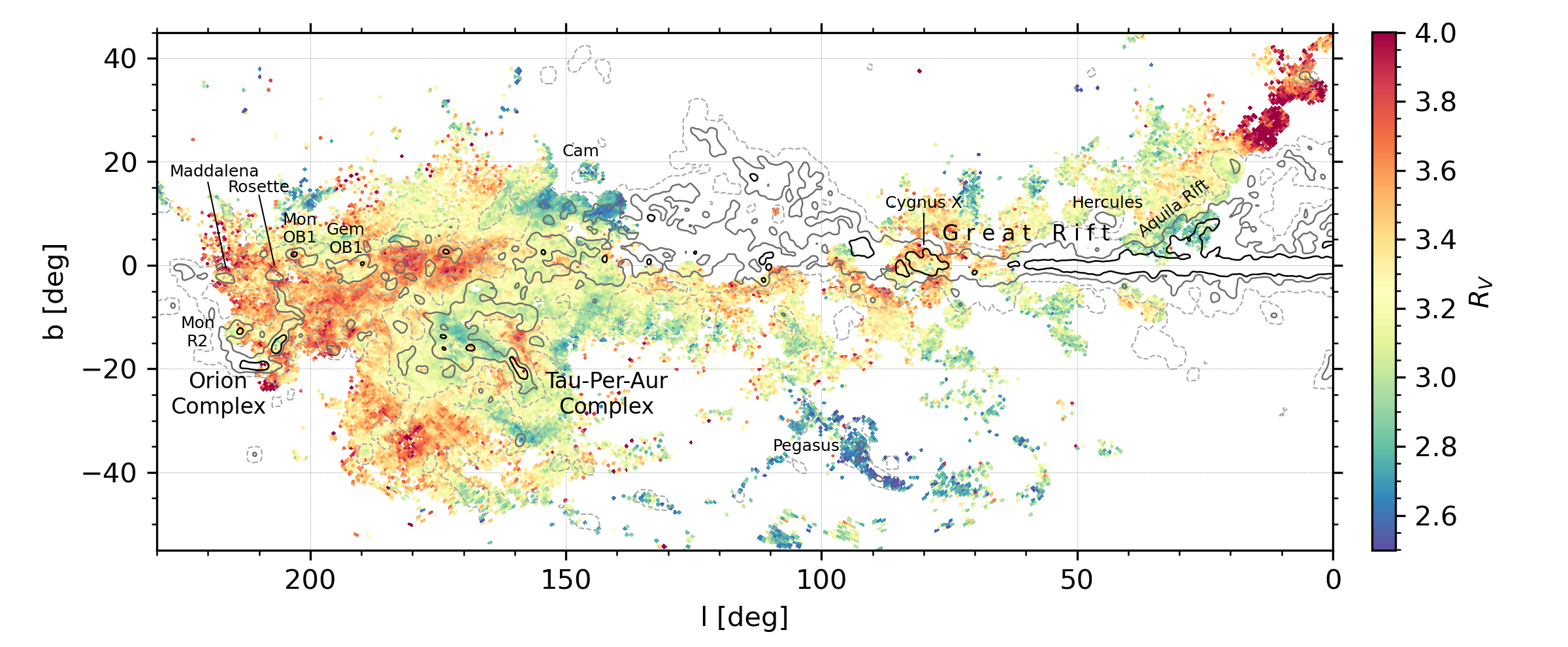}
    \includegraphics[width=\linewidth]{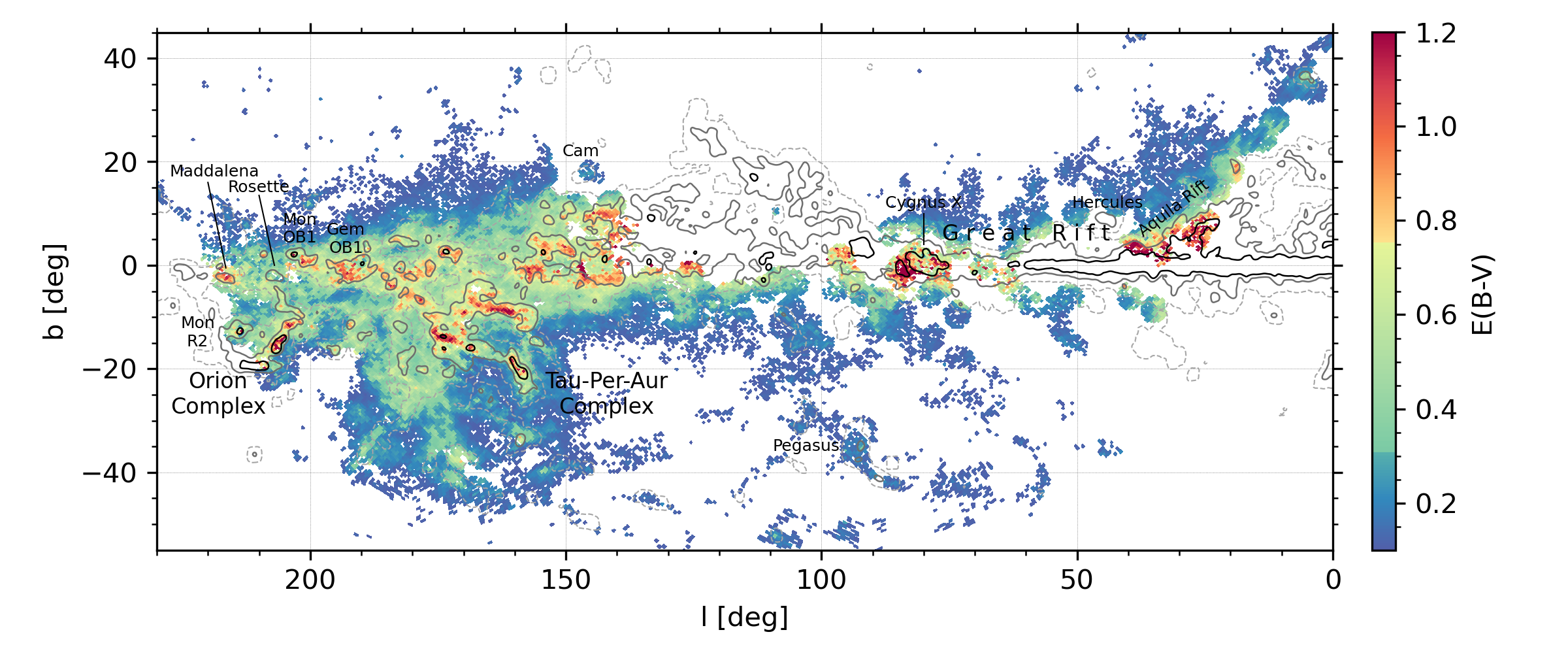}
    \caption{Maps of $\RV$ (upper panel) and $\ebv$ (bottom panel) for the Galactic disk. Median values of $\RV$ or $\ebv$ in each direction are represented by colored dots. The gray contours show the smooth CO map from \citet{2016A&A...594A..10P}, which become increasingly intense (from dashed gray to solid black) as intensity increases. White areas are regions unobserved or non-reliable. }
    \label{fig:disk}
\end{figure*}

In Fig.~\ref{fig:disk}, we present a comparison of the distribution of $\RV$ and $\ebv$ for reliable sightlines in the Galactic disk. To trace molecular clouds, we smooth the J=1--0 CO velocity-integrated emission maps (type 2) produced by \citet{2016A&A...594A..10P}. Several nearby molecular clouds are marked in the Figure. A striking correlation can be observed between the distribution of $\RV$ and the molecular clouds traced by CO. Specifically, the $\RV$ values within the regions of molecular clouds are consistently lower than those in the surrounding regions. While the number of stars towards the cores of clouds decreases due to high extinction, our analysis suggests that the $\RV$ differences inside and outside the clouds are not due to the differences in detection depth. This pattern is observed in almost all clouds, but is most evident in the Galactic anti-center direction, where the data are most abundant. This finding suggests that molecular clouds play a crucial role in the chemical and size evolution of dust grains. 

The mean $\RV$ values within each cloud vary and are not dependent on the extinction. Specifically, the molecular clouds Cam, Aquila Rift, and Pegasus have very low $\RV$ values of 2.6 -- 2.8, while Taurus, California (in the Tau-Per-Aur complex), and Mon R2 have moderate $\RV$ values of about 3.0. Additionally, Maddalena, Mon OB1, Gem OB1, Orion Complex, Cygnus X, and others have high $\RV$ values, averaging around 3.3. More detailed explorations on how $\RV$ varies between different molecular clouds will be presented in a future paper.

The phenomenon of lower $\RV$ value inside the molecular clouds may be related to different types of dust environments.
\citet{2006ARA&A..44..367S} classified interstellar clouds into four types based on their optical depth, as follows:
\begin{itemize}
\item Diffuse atomic clouds with $\AV < 0.2$, which are fully exposed to the interstellar radiation field.
\item Diffuse molecular clouds with $\AV < 1$, which have a weaker radiation field that allows for a significant fraction of molecular hydrogen to form locally.
\item Translucent clouds with $1< \AV < 5$, which are optically thick clouds where photoprocessing still plays a crucial role in the overall chemistry. The chemistry in this regime differs significantly from that of diffuse molecular clouds because of the reduced electron fraction and the prevalence of molecular carbon in the form of CO, as noted in \citep{2006ARA&A..44..367S}.
\item Dense molecular clouds with $\AV > 5$. Since few stars in our sample have $\AV > 5$,  such clouds are not discussed in this paper.
\end{itemize}
However, this classification scheme oversimplifies the situation in our study, as it does not consider scenarios where multiple diffuse clouds contribute to total extinction along a sightline. Additionally, when a sightline direction passes through a translucent cloud, it often encounters diffuse clouds first. Therefore, a more precise classification would require considering local extinction gradients instead of the integral extinction of the entire dust column.

Based on the present findings, it appears that the size of dust grains in translucent clouds is influenced by both growth and reduction physical processes. Coagulation and accretion processes contribute to the growth of dust grain size, while photodissociation and collision processes lead to their reduction. These opposing factors balance each other out, resulting in a decrease in the average dust grain size within the extinction range of about $0.3 < \ebv < 1.2$. Consequently, this decrease in grain size leads to a decrease in the $\RV$ value within the molecular cloud.
On the other hand, studies that have investigated samples with larger extinction, such as \citet{2013MNRAS.428.1606F}, have shown that $\RV$ values increase with increasing extinction. 
This difference in behavior could be attributed to the dominance of condensation and accretion processes in regions with higher extinction. 
A very clear positive correlation between Rv and extinction was not seen in this study, possibly due to the lack of samples with larger extinction and thus the inability to detect regions of dense clouds or nuclei.

\subsection{Correlation between $\RV$ and extinction}
The alignment of the $\RV$ distribution with molecular clouds prompts us to investigate the correlation between $\RV$ and extinction. Unlike $\ebv$ (or $\AV$), which accounts for the cumulative reddening (or extinction) effects of dust grains along the line of sight, $\RV$ only reflects the average properties of the dust grains along the same path. As the length of the dust column can vary, the impact of local dust grains on $\RV$ and $\ebv$ (or $\AV$) also differs. Therefore, we exclude stars with a Galactic vertical distance $\vert Z \vert <$ 200\,pc, allowing us to penetrate the dust disk. However, we would like to stress that a comprehensive understanding of the relationship between $\RV$ and extinction requires further investigation using 3D $\RV$ and extinction maps in future studies.

\begin{figure*}
    \centering
    \includegraphics[width=\linewidth]{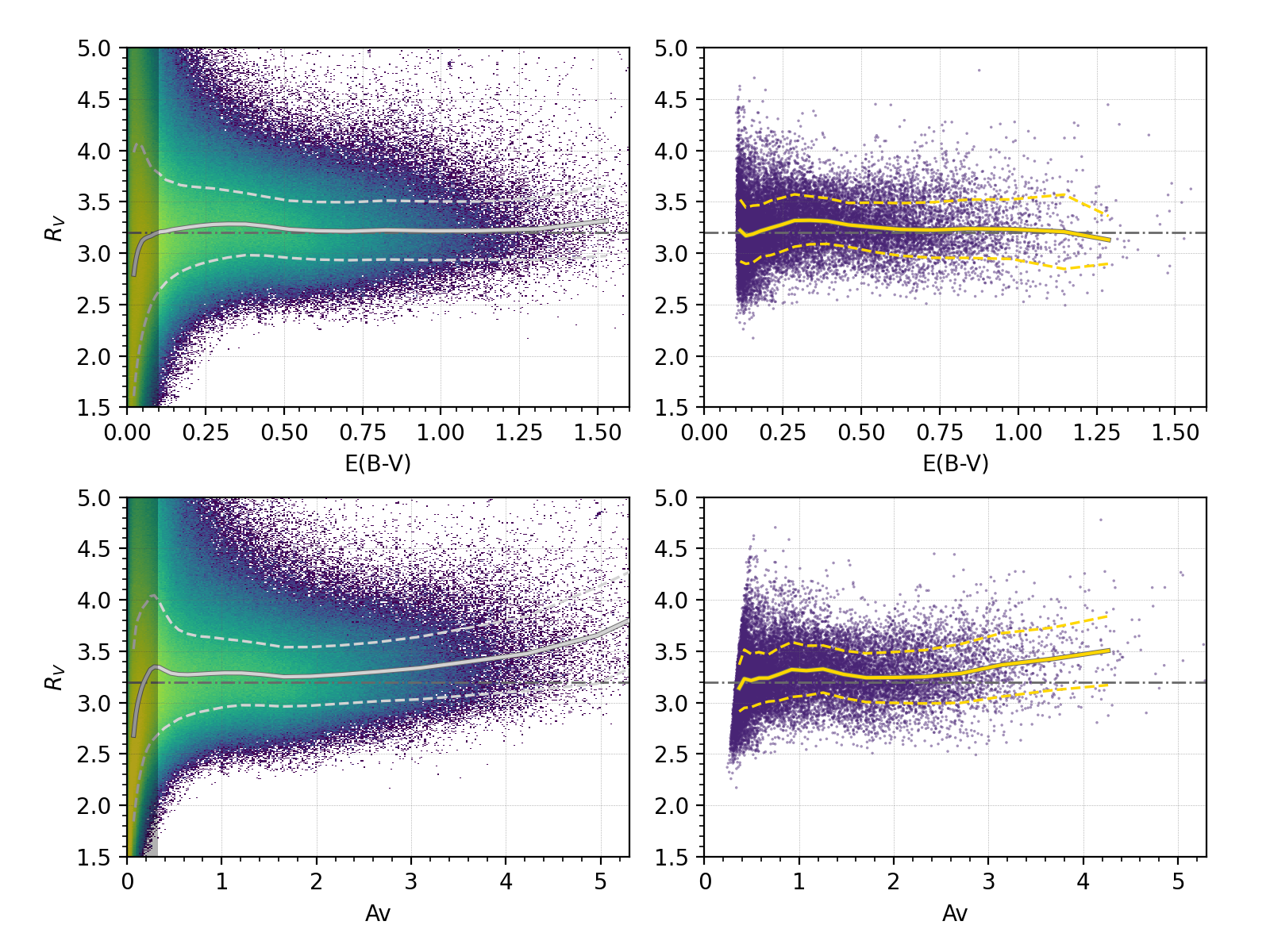}
    \caption{Top panel: $\RV$-$\ebv$ diagram for the reliable stars with $|Z|>200$\,pc (left panel) and the corresponding sightlines (right panel). Bottom panel: Same as the top panel but for $\RV$ - $\AV$ diagram. as the horizontal coordinate. Solid lines represent the medians of binned points, while the dashed lines depict the 1$\sigma$ region. The dot-dashed lines marking $\RV$ = 3.2 are plotted to guide eyes.
    }
    \label{fig:EBV}
\end{figure*}

In Fig.~\ref{fig:EBV}, we show the correlations between $\RV$ and extinction ($\ebv$ and $\AV$) for the reliable stars with $\vert Z \vert > 200$\,pc, as well as the corresponding sightlines. In general, the distribution of $\RV$ for the reliable stars and sightlines are both independent of $\ebv$, which is similar to that of \citet{2016ApJ...821...78S} (see their Fig.~16).
The correlations between $\RV$ and $\AV$ exhibit a pattern similar to that observed for $\ebv$, with a notable exception of a positive relationship emerging at $\AV > 2.5$.
Note that the $\RV$-$\ebv$ relation does not obviously exhibit the trend described in Section~\ref{sec:clouds}, i.e., the $\RV$ values within the regions of molecular clouds are lower than those in the surrounding regions.
This is because the $\RV$ and $\ebv$ of individual clouds vary greatly, thus smoothing out this effect in the overall statistics.
For example, in our current studies of the Orion complex and the Tau-Per-Aur complex, we have observed a correlation between $\RV$ and $\ebv$ within individual clouds. 
Specifically, for stars with $\ebv <  1.5$, we have found an inverse correlation between $\RV$ and $\ebv$ in the Orion complex, while a positive correlation in the Tau-Per-Aur complex. 
Further investigations and a more detailed study will be conducted in our future work.

\subsection{Correlation between $\RV$ and other parameters of interstellar clouds} \label{sec:relation}

\begin{figure*}
    \centering
    \includegraphics[width=\linewidth]{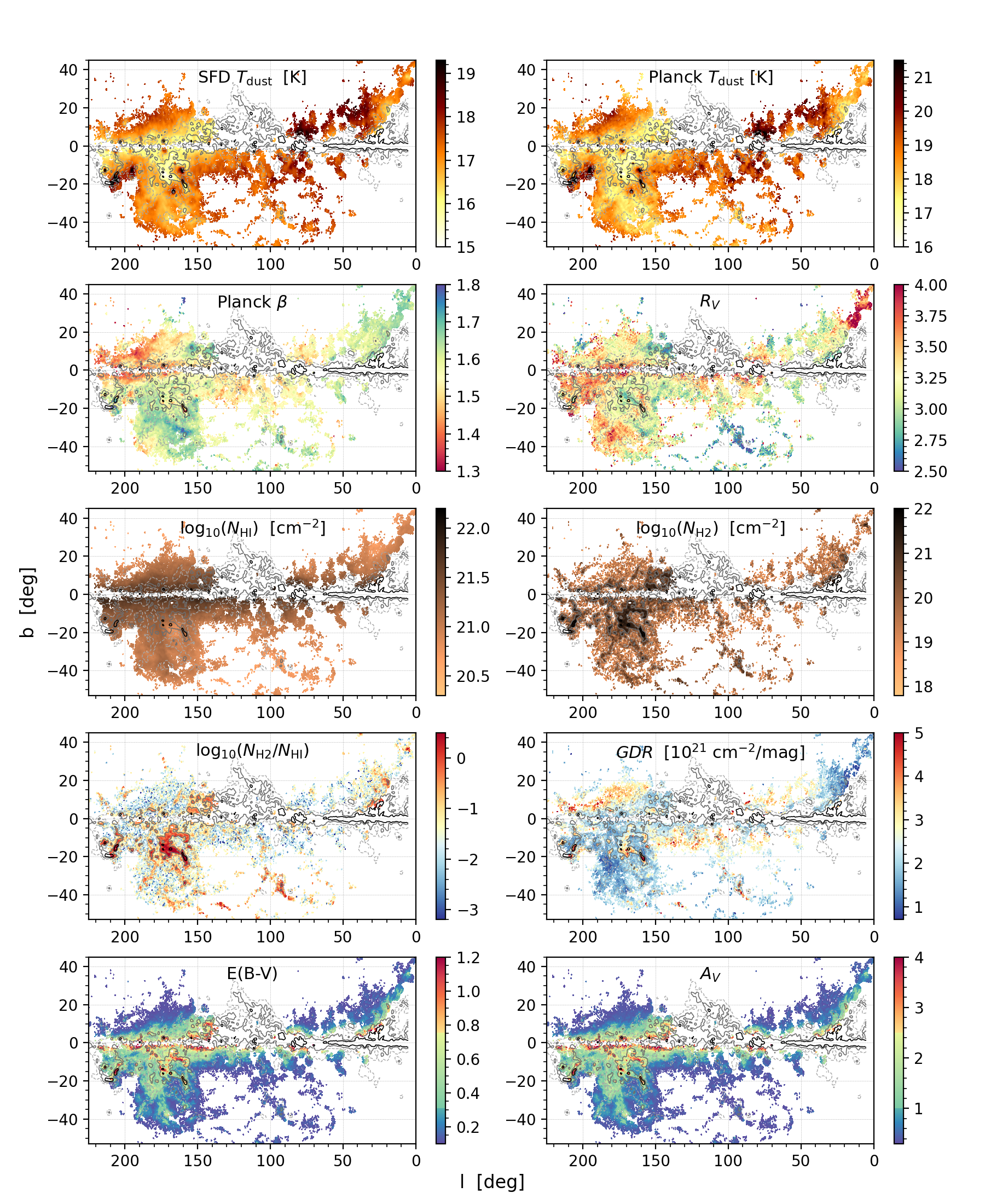}
    \caption{Spatial distribution of various interstellar medium parameters. Moving from the top left to the bottom right panels, the colored dots represent the SFD $T_{\rm dust}$, $Planck$ $T_{\rm dust}$, $Planck$ $\beta$, $\RV$, $N_{\rm HI}$, $N_{\rm H_2}$, $N_{\rm H_2}/N_{\rm HI}$, $GDR$, $\ebv$, and $\AV$, respectively. Each dot corresponds to the median value of a reliable sightline. The gray contours show the smooth CO map from \citet{2016A&A...594A..10P}, which become increasingly intense (from dashed gray to solid black) as intensity increases. White areas are regions unobserved or non-reliable.}
    \label{fig:Correlation_space}
\end{figure*}

In this subsection, we investigate the correlation between $\RV$ and other parameters of interstellar clouds, as follows:
\begin{enumerate}
\item[(a)] Dust temperature T$_{\rm dust}$ from SFD;
\item[(b)] Dust temperature T$_{\rm dust}$ and dust emissivity spectral index $\beta$ from \citet{2019A&A...623A..21I};
\item[(c)] Neutral atomic hydrogen column density $N_{\rm HI}$ from \citet{2019A&A...623A..21I};
\item[(d)] Molecular hydrogen column density $N_{\rm H_2}$ derived from the CO ($J$ = 1 -- 0, type 2) integrated line intensity $W_{\rm CO}$ from \citet{2016A&A...594A..10P}, by using a CO-to-${\rm H_2}$ conversion factor of $2 \times 10^{20} \rm{cm^{-2} (K\,km\,s^{-1})^{-1}}$ \citep{2013ARA&A..51..207B};
\item[(e)]  Ratio between $N_{\rm H_2}$ and $N_{\rm HI}$ using (c) and (d);
\item[(f)]  The gas-dust-ratio ($GDR$) using our resulting $\AV$ values, (c), and (d), defined as $GDR = (N_{\rm HI} + 2N_{\rm H_2})/ \AV$.
\end{enumerate}    
As shown in Fig.~\ref{fig:Correlation_space}, we smooth these maps to match the resolution of our $\RV$ map, with the exception of the SFD T$_{\rm dust}$ map, which has a lower resolution than our map. We note that the aforementioned data represent integrals over infinite distances, while our $\RV$, $\ebv$, and $\AV$ measurements only pertain to the column space between the observer and the star. To minimize potential errors, we only included sources outside the dust disk ($\vert Z \vert \ge 200$\,pc) in the subsequent analysis. 

\begin{figure*}
    \centering
    \includegraphics[width=\linewidth]{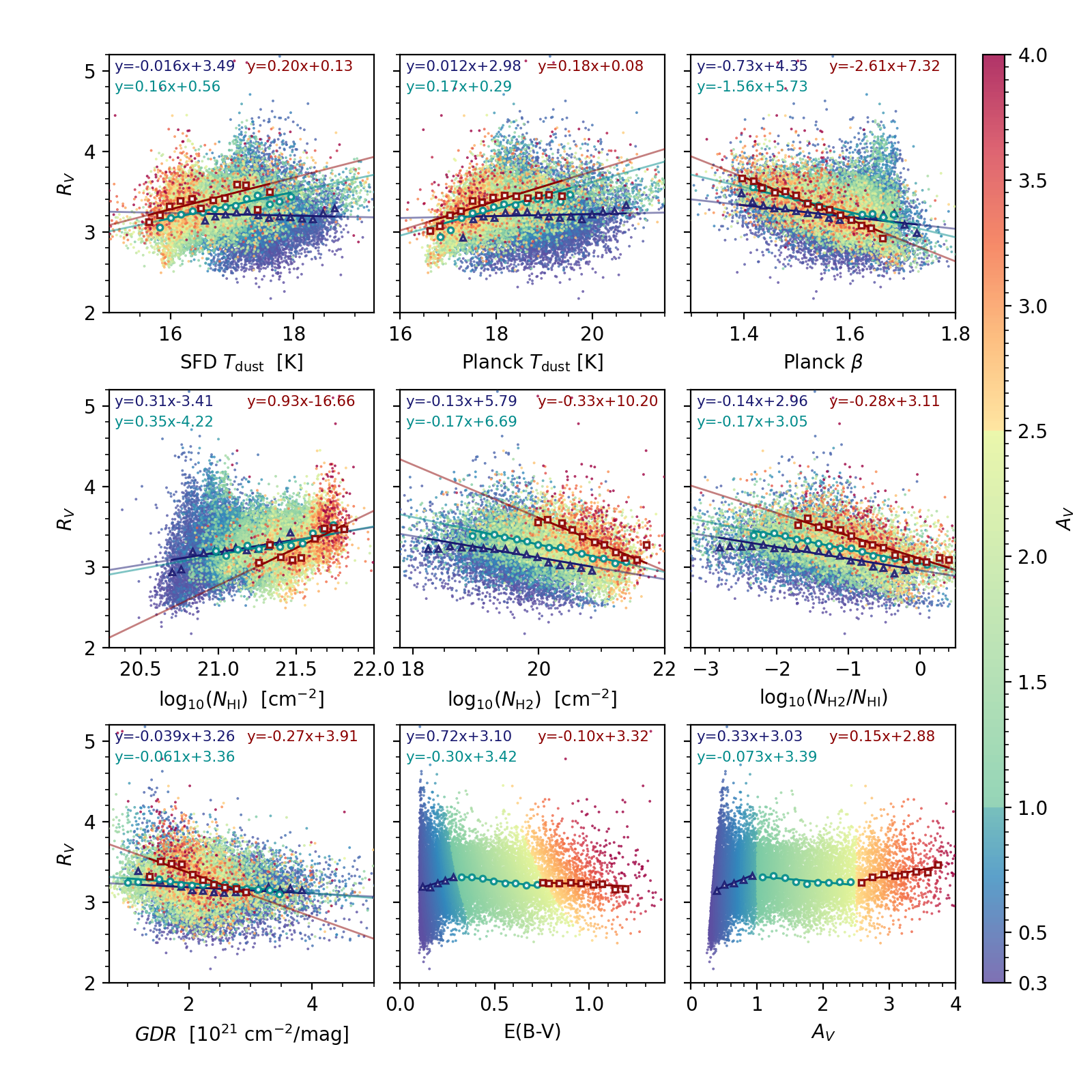}
    \caption{The correlation between $\RV$ and various interstellar medium parameters. Moving from the top left to the bottom right panel, the horizontal axis represents the SFD $T_{\rm dust}$, $Planck$ $T_{\rm dust}$, $Planck$ $\beta$, $N_{\rm HI}$, $N_{\rm H_2}$, $N_{\rm H_2}/N_{\rm HI}$, $GDR$, $\ebv$, and $\AV$, respectively. Each dot on the plot corresponds to a reliable sightline, with the color of the dot indicating the $\AV$ value. The median binned points for the subsets with $0.32< \AV<1$, $1<\AV<2.5$, and $\AV > 2.5$ are represented by blue, cyan, and red, respectively. The curves and equations obtained from the linear fits are also shown and labeled in each panel.}
    \label{fig:Correlation}
\end{figure*}

In Fig.~\ref{fig:Correlation}, we present the correlations between $\RV$ and the aforementioned parameters for each reliable sightline. The sightlines are categorized into three groups based on their $\AV$ values. These groups correspond to the following $\AV$ ranges: $0.32<\AV<1$ for diffuse molecular clouds, $1<\AV<2.5$ for low-extinction translucent molecular clouds, and $\AV > 2.5$ for high-extinction translucent molecular clouds. Weighted linear fits are performed on the median values of the parameters for each subset, and the resulting fits are displayed in the figure as well.

We observe a similar pattern in the correlation between $\RV$ and $T_{\rm dust}$ for both the SFD and $Planck$ results. The low- and high-extinction translucent molecular clouds exhibit lower $T_{\rm dust}$ values and show a positive correlation between $\RV$ and $T_{\rm dust}$. On the other hand, the diffuse molecular clouds have relatively higher $T_{\rm dust}$ values and demonstrate a negative correlation between $\RV$ and $T_{\rm dust}$.

We find a strong negative correlation between $\beta$ and $\RV$ in both the translucent and diffuse molecular clouds, which is consistent with the results reported by \citet{2016ApJ...821...78S}. This indicates that high $\beta$ values, which indicate steep far-infrared emission spectra, are associated with steep extinction curves at low $\RV$. It is worth noting that although the range of $\beta$ is the same in both types of clouds, the anti-correlation is much weaker in diffuse molecular clouds.

We observe a consistent positive correlation between $\RV$ and $N_{\rm HI}$ in all types of clouds. Conversely, negative correlations are found between $\RV$ and $N_{\rm H_2}$, as well as between $\RV$ and $N_{\rm H_2}/N_{\rm HI}$ across all cloud types. Furthermore, we find a negative correlation between $\RV$ and $GDR$ in all types of clouds. These relationships exhibit a stronger correlation as the extinction increases.
Although not within the scope of this study, these findings provide important clues for future discussions on dust properties in various interstellar environments.

\section{Summary} \label{sec:Summary}

Using high-precision CERs between colors from the optical to the mid-IR, we have measured the $\RV$ values of about 3 million LAMOST stars in our study. To account for the combined impact of stellar SED and extinction, we have developed a robust forward modeling approach based on a model constructed using BOSZ stellar spectra and the F99 extinction curve. To validate our results, we compared our derived $\RV$ values with the literature and found good agreement. We have divided the sample stars into different sightlines and calculated the median $\RV$ value for each direction. This allowed us to create a 2D $\RV$ map within the LAMOST footprint, spanning from the Galactic disk to the Galactic halo. Based on the analysis of reliable sightlines, we summarize our findings as follows:
\begin{enumerate}
\item Overall, the distribution of $\RV$ is well-described by a Gaussian distribution with a mean of 3.25 and a dispersion of 0.25, with a slight excess at the high end. The variability of $\RV$ within the Galactic disk exhibits a wide range, manifesting at various scales from small structures within individual molecular clouds to larger scales spanning kiloparsecs.
\item  The spatial distribution of $\RV$ closely aligns with the shape of molecular clouds. Specifically, we observe lower $\RV$ values within the interior regions of molecular clouds compared to their surrounding areas. Although the average $\RV$ may vary across different molecular clouds, this coincidence suggests that molecular clouds play a crucial role in the chemical and size evolution of dust grains.
\item In the $\ebv$ interval ranging from 0.1 to 1.25, we find that $\RV$ is largely independent of extinction. Additionally, we have investigated the correlations between $\RV$ and other interstellar parameters, such as $T_{\rm dust}$, $\beta$, $N_{\rm HI}$, $N_{\rm H_2}$, $N_{\rm H_2}/N_{\rm HI}$, and $GDR$. Notably, we observe that these relationships vary with the level of extinction, providing valuable insights into the diverse properties of dust in different interstellar environments.
\end{enumerate}

In addition to giving us further insight into the physics of the dust, the $\RV$ map also helps us to do precision extinction correction.   
After correcting the all-sky two-dimensional (2D) reddening map \citep{2022ApJS..260...17S} and investigating the temperature and extinction dependence of reddening coefficients \citep{2023ApJS..264...14Z},
in this study, we also laid foundation for addressing the third challenge outlined in ZY23, i.e., the significant impact of the spatial variation of $\RV$ on extinction corrections. In the future, we will synthesize factors to create an accurate extinction correction toolkit.

\begin{acknowledgments}

We acknowledge the anonymous referee for his or her valuable comments to improve the clarity and quality of the manuscript.
We acknowledge Prof. Biwei Jiang for her useful discussions.
This work is supported by the National Key Basic R\&D Program of China via 2019YFA0405500 and the National Natural Science Foundation of China through the projects NSFC 12222301, 12173007, and 12173034.
We acknowledge the science research grants from the China Manned Space Project with NO. CMS-CSST-2021-A08, CMS-CSST-2021-A09 and and CMS-CSST-2021-B03. 

This work has made use of data products from the LAMOST, GALEX, PS1, $\gaia$, SDSS, 2MASS, and WISE. 
Guoshoujing Telescope (the Large Sky Area Multi-Object Fiber Spectroscopic Telescope, LAMOST) is a National Major Scientific Project built by the Chinese Academy of Sciences. 
Funding for the project has been provided by the National Development and Reform Commission. 
LAMOST is operated and managed by the National Astronomical Observatories, Chinese Academy of Sciences. 

\end{acknowledgments}

%



\bibliography{main}{}

\begin{thebibliography}{}
\expandafter\ifx\csname natexlab\endcsname\relax\def\natexlab#1{#1}\fi
\providecommand{\url}[1]{\href{#1}{#1}}
\providecommand{\dodoi}[1]{doi:~\href{http://doi.org/#1}{\nolinkurl{#1}}}
\providecommand{\doeprint}[1]{\href{http://ascl.net/#1}{\nolinkurl{http://ascl.net/#1}}}
\providecommand{\doarXiv}[1]{\href{https://arxiv.org/abs/#1}{\nolinkurl{https://arxiv.org/abs/#1}}}

\bibitem[{{Alam} {et~al.}(2015){Alam}, {Albareti}, {Allende Prieto}, {Anders},
  {Anderson}, {Anderton}, {Andrews}, {Armengaud}, {Aubourg}, {Bailey}, {Basu},
  {Bautista}, {Beaton}, {Beers}, {Bender}, {Berlind}, {Beutler}, {Bhardwaj},
  {Bird}, {Bizyaev}, {Blake}, {Blanton}, {Blomqvist}, {Bochanski}, {Bolton},
  {Bovy}, {Shelden Bradley}, {Brandt}, {Brauer}, {Brinkmann}, {Brown},
  {Brownstein}, {Burden}, {Burtin}, {Busca}, {Cai}, {Capozzi}, {Carnero
  Rosell}, {Carr}, {Carrera}, {Chambers}, {Chaplin}, {Chen}, {Chiappini},
  {Chojnowski}, {Chuang}, {Clerc}, {Comparat}, {Covey}, {Croft}, {Cuesta},
  {Cunha}, {da Costa}, {Da Rio}, {Davenport}, {Dawson}, {De Lee}, {Delubac},
  {Deshpande}, {Dhital}, {Dutra-Ferreira}, {Dwelly}, {Ealet}, {Ebelke},
  {Edmondson}, {Eisenstein}, {Ellsworth}, {Elsworth}, {Epstein}, {Eracleous},
  {Escoffier}, {Esposito}, {Evans}, {Fan}, {Fern{\'a}ndez-Alvar}, {Feuillet},
  {Filiz Ak}, {Finley}, {Finoguenov}, {Flaherty}, {Fleming}, {Font-Ribera},
  {Foster}, {Frinchaboy}, {Galbraith-Frew}, {Garc{\'\i}a},
  {Garc{\'\i}a-Hern{\'a}ndez}, {Garc{\'\i}a P{\'e}rez}, {Gaulme}, {Ge},
  {G{\'e}nova-Santos}, {Georgakakis}, {Ghezzi}, {Gillespie}, {Girardi},
  {Goddard}, {Gontcho}, {Gonz{\'a}lez Hern{\'a}ndez}, {Grebel}, {Green},
  {Grieb}, {Grieves}, {Gunn}, {Guo}, {Harding}, {Hasselquist}, {Hawley},
  {Hayden}, {Hearty}, {Hekker}, {Ho}, {Hogg}, {Holley-Bockelmann}, {Holtzman},
  {Honscheid}, {Huber}, {Huehnerhoff}, {Ivans}, {Jiang}, {Johnson},
  {Kinemuchi}, {Kirkby}, {Kitaura}, {Klaene}, {Knapp}, {Kneib}, {Koenig},
  {Lam}, {Lan}, {Lang}, {Laurent}, {Le Goff}, {Leauthaud}, {Lee}, {Lee},
  {Licquia}, {Liu}, {Long}, {L{\'o}pez-Corredoira}, {Lorenzo-Oliveira},
  {Lucatello}, {Lundgren}, {Lupton}, {Mack}, {Mahadevan}, {Maia}, {Majewski},
  {Malanushenko}, {Malanushenko}, {Manchado}, {Manera}, {Mao}, {Maraston},
  {Marchwinski}, {Margala}, {Martell}, {Martig}, {Masters}, {Mathur},
  {McBride}, {McGehee}, {McGreer}, {McMahon}, {M{\'e}nard}, {Menzel},
  {Merloni}, {M{\'e}sz{\'a}ros}, {Miller}, {Miralda-Escud{\'e}}, {Miyatake},
  {Montero-Dorta}, {More}, {Morganson}, {Morice-Atkinson}, {Morrison},
  {Mosser}, {Muna}, {Myers}, {Nandra}, {Newman}, {Neyrinck}, {Nguyen},
  {Nichol}, {Nidever}, {Noterdaeme}, {Nuza}, {O'Connell}, {O'Connell},
  {O'Connell}, {Ogando}, {Olmstead}, {Oravetz}, {Oravetz}, {Osumi}, {Owen},
  {Padgett}, {Padmanabhan}, {Paegert}, {Palanque-Delabrouille}, {Pan},
  {Parejko}, {P{\^a}ris}, {Park}, {Pattarakijwanich}, {Pellejero-Ibanez},
  {Pepper}, {Percival}, {P{\'e}rez-Fournon}, {P{\'e}rez-R{\`a}fols},
  {Petitjean}, {Pieri}, {Pinsonneault}, {Porto de Mello}, {Prada}, {Prakash},
  {Price-Whelan}, {Protopapas}, {Raddick}, {Rahman}, {Reid}, {Rich}, {Rix},
  {Robin}, {Rockosi}, {Rodrigues}, {Rodr{\'\i}guez-Torres}, {Roe}, {Ross},
  {Ross}, {Rossi}, {Ruan}, {Rubi{\~n}o-Mart{\'\i}n}, {Rykoff},
  {Salazar-Albornoz}, {Salvato}, {Samushia}, {S{\'a}nchez}, {Santiago},
  {Sayres}, {Schiavon}, {Schlegel}, {Schmidt}, {Schneider}, {Schultheis},
  {Schwope}, {Sc{\'o}ccola}, {Scott}, {Sellgren}, {Seo}, {Serenelli}, {Shane},
  {Shen}, {Shetrone}, {Shu}, {Silva Aguirre}, {Sivarani}, {Skrutskie},
  {Slosar}, {Smith}, {Sobreira}, {Souto}, {Stassun}, {Steinmetz}, {Stello},
  {Strauss}, {Streblyanska}, {Suzuki}, {Swanson}, {Tan}, {Tayar}, {Terrien},
  {Thakar}, {Thomas}, {Thomas}, {Thompson}, {Tinker}, {Tojeiro}, {Troup},
  {Vargas-Maga{\~n}a}, {Vazquez}, {Verde}, {Viel}, {Vogt}, {Wake}, {Wang},
  {Weaver}, {Weinberg}, {Weiner}, {White}, {Wilson}, {Wisniewski},
  {Wood-Vasey}, {Ye`che}, {York}, {Zakamska}, {Zamora}, {Zasowski}, {Zehavi},
  {Zhao}, {Zheng}, {Zhou}, {Zhou}, {Zou}, \& {Zhu}}]{2015ApJS..219...12A}
{Alam}, S., {Albareti}, F.~D., {Allende Prieto}, C., {et~al.} 2015, \apjs, 219,
  12, \dodoi{10.1088/0067-0049/219/1/12}

\bibitem[{{Bohlin} {et~al.}(2017){Bohlin}, {M{\'e}sz{\'a}ros}, {Fleming},
  {Gordon}, {Koekemoer}, \& {Kov{\'a}cs}}]{2017AJ....153..234B}
{Bohlin}, R.~C., {M{\'e}sz{\'a}ros}, S., {Fleming}, S.~W., {et~al.} 2017, \aj,
  153, 234, \dodoi{10.3847/1538-3881/aa6ba9}

\bibitem[{{Bolatto} {et~al.}(2013){Bolatto}, {Wolfire}, \&
  {Leroy}}]{2013ARA&A..51..207B}
{Bolatto}, A.~D., {Wolfire}, M., \& {Leroy}, A.~K. 2013, \araa, 51, 207,
  \dodoi{10.1146/annurev-astro-082812-140944}

\bibitem[{{Cardelli} {et~al.}(1989){Cardelli}, {Clayton}, \&
  {Mathis}}]{1989ApJ...345..245C}
{Cardelli}, J.~A., {Clayton}, G.~C., \& {Mathis}, J.~S. 1989, \apj, 345, 245,
  \dodoi{10.1086/167900}

\bibitem[{{Chambers} \& {Pan-STARRS Team}(2018)}]{2018AAS...23110201C}
{Chambers}, K., \& {Pan-STARRS Team}. 2018, in American Astronomical Society
  Meeting Abstracts, Vol. 231, American Astronomical Society Meeting Abstracts
  \#231, 102.01

\bibitem[{{Cui} {et~al.}(2012){Cui}, {Zhao}, {Chu}, {Li}, {Li}, {Zhang}, {Su},
  {Yao}, {Wang}, {Xing}, {Li}, {Zhu}, {Wang}, {Gu}, {Luo}, {Xu}, {Zhang},
  {Liu}, {Zhang}, {Yang}, {Cao}, {Chen}, {Chen}, {Chen}, {Chen}, {Chu}, {Feng},
  {Gong}, {Hou}, {Hu}, {Hu}, {Hu}, {Jia}, {Jiang}, {Jiang}, {Jiang}, {Jin},
  {Li}, {Li}, {Li}, {Liu}, {Liu}, {Lu}, {Mao}, {Men}, {Qi}, {Qi}, {Shi},
  {Tang}, {Tao}, {Wang}, {Wang}, {Wang}, {Wang}, {Wang}, {Wang}, {Wang},
  {Wang}, {Wang}, {Wang}, {Wang}, {Wang}, {Xu}, {Xu}, {Yang}, {Yu}, {Yuan},
  {Yuan}, {Zhai}, {Zhang}, {Zhang}, {Zhang}, {Zhao}, {Zhou}, {Zhou}, {Zhu}, \&
  {Zou}}]{2012RAA....12.1197C}
{Cui}, X.-Q., {Zhao}, Y.-H., {Chu}, Y.-Q., {et~al.} 2012, Research in Astronomy
  and Astrophysics, 12, 1197, \dodoi{10.1088/1674-4527/12/9/003}

\bibitem[{{Draine}(2003)}]{2003ARA&A..41..241D}
{Draine}, B.~T. 2003, \araa, 41, 241,
  \dodoi{10.1146/annurev.astro.41.011802.094840}

\bibitem[{{Fitzpatrick}(1999)}]{1999PASP..111...63F}
{Fitzpatrick}, E.~L. 1999, \pasp, 111, 63, \dodoi{10.1086/316293}

\bibitem[{{Fitzpatrick} \& {Massa}(1986)}]{1986ApJ...307..286F}
{Fitzpatrick}, E.~L., \& {Massa}, D. 1986, \apj, 307, 286,
  \dodoi{10.1086/164415}

\bibitem[{{Fitzpatrick} \& {Massa}(1988)}]{1988ApJ...328..734F}
---. 1988, \apj, 328, 734, \dodoi{10.1086/166332}

\bibitem[{{Fitzpatrick} \& {Massa}(1990)}]{1990ApJS...72..163F}
---. 1990, \apjs, 72, 163, \dodoi{10.1086/191413}

\bibitem[{{Fitzpatrick} \& {Massa}(2007)}]{2007ApJ...663..320F}
---. 2007, \apj, 663, 320, \dodoi{10.1086/518158}

\bibitem[{{Fitzpatrick} {et~al.}(2019){Fitzpatrick}, {Massa}, {Gordon},
  {Bohlin}, \& {Clayton}}]{2019ApJ...886..108F}
{Fitzpatrick}, E.~L., {Massa}, D., {Gordon}, K.~D., {Bohlin}, R., \& {Clayton},
  G.~C. 2019, \apj, 886, 108, \dodoi{10.3847/1538-4357/ab4c3a}

\bibitem[{{Foster} {et~al.}(2013){Foster}, {Mandel}, {Pineda}, {Covey}, {Arce},
  \& {Goodman}}]{2013MNRAS.428.1606F}
{Foster}, J.~B., {Mandel}, K.~S., {Pineda}, J.~E., {et~al.} 2013, \mnras, 428,
  1606, \dodoi{10.1093/mnras/sts144}

\bibitem[{{Gaia Collaboration} {et~al.}(2021){Gaia Collaboration}, {Brown},
  {Vallenari}, {Prusti}, {de Bruijne}, {Babusiaux}, {Biermann}, {Creevey},
  {Evans}, {Eyer}, {Hutton}, {Jansen}, {Jordi}, {Klioner}, {Lammers},
  {Lindegren}, {Luri}, {Mignard}, {Panem}, {Pourbaix}, {Randich}, {Sartoretti},
  {Soubiran}, {Walton}, {Arenou}, {Bailer-Jones}, {Bastian}, {Cropper},
  {Drimmel}, {Katz}, {Lattanzi}, {van Leeuwen}, {Bakker}, {Cacciari},
  {Casta{\~n}eda}, {De Angeli}, {Ducourant}, {Fabricius}, {Fouesneau},
  {Fr{\'e}mat}, {Guerra}, {Guerrier}, {Guiraud}, {Jean-Antoine Piccolo},
  {Masana}, {Messineo}, {Mowlavi}, {Nicolas}, {Nienartowicz}, {Pailler},
  {Panuzzo}, {Riclet}, {Roux}, {Seabroke}, {Sordo}, {Tanga}, {Th{\'e}venin},
  {Gracia-Abril}, {Portell}, {Teyssier}, {Altmann}, {Andrae}, {Bellas-Velidis},
  {Benson}, {Berthier}, {Blomme}, {Brugaletta}, {Burgess}, {Busso}, {Carry},
  {Cellino}, {Cheek}, {Clementini}, {Damerdji}, {Davidson}, {Delchambre},
  {Dell'Oro}, {Fern{\'a}ndez-Hern{\'a}ndez}, {Galluccio}, {Garc{\'\i}a-Lario},
  {Garcia-Reinaldos}, {Gonz{\'a}lez-N{\'u}{\~n}ez}, {Gosset}, {Haigron},
  {Halbwachs}, {Hambly}, {Harrison}, {Hatzidimitriou}, {Heiter},
  {Hern{\'a}ndez}, {Hestroffer}, {Hodgkin}, {Holl}, {Jan{\ss}en}, {Jevardat de
  Fombelle}, {Jordan}, {Krone-Martins}, {Lanzafame}, {L{\"o}ffler}, {Lorca},
  {Manteiga}, {Marchal}, {Marrese}, {Moitinho}, {Mora}, {Muinonen}, {Osborne},
  {Pancino}, {Pauwels}, {Petit}, {Recio-Blanco}, {Richards}, {Riello},
  {Rimoldini}, {Robin}, {Roegiers}, {Rybizki}, {Sarro}, {Siopis}, {Smith},
  {Sozzetti}, {Ulla}, {Utrilla}, {van Leeuwen}, {van Reeven}, {Abbas}, {Abreu
  Aramburu}, {Accart}, {Aerts}, {Aguado}, {Ajaj}, {Altavilla}, {{\'A}lvarez},
  {{\'A}lvarez Cid-Fuentes}, {Alves}, {Anderson}, {Anglada Varela}, {Antoja},
  {Audard}, {Baines}, {Baker}, {Balaguer-N{\'u}{\~n}ez}, {Balbinot}, {Balog},
  {Barache}, {Barbato}, {Barros}, {Barstow}, {Bartolom{\'e}}, {Bassilana},
  {Bauchet}, {Baudesson-Stella}, {Becciani}, {Bellazzini}, {Bernet}, {Bertone},
  {Bianchi}, {Blanco-Cuaresma}, {Boch}, {Bombrun}, {Bossini}, {Bouquillon},
  {Bragaglia}, {Bramante}, {Breedt}, {Bressan}, {Brouillet}, {Bucciarelli},
  {Burlacu}, {Busonero}, {Butkevich}, {Buzzi}, {Caffau}, {Cancelliere},
  {C{\'a}novas}, {Cantat-Gaudin}, {Carballo}, {Carlucci}, {Carnerero},
  {Carrasco}, {Casamiquela}, {Castellani}, {Castro-Ginard}, {Castro Sampol},
  {Chaoul}, {Charlot}, {Chemin}, {Chiavassa}, {Cioni}, {Comoretto}, {Cooper},
  {Cornez}, {Cowell}, {Crifo}, {Crosta}, {Crowley}, {Dafonte}, {Dapergolas},
  {David}, {David}, {de Laverny}, {De Luise}, {De March}, {De Ridder}, {de
  Souza}, {de Teodoro}, {de Torres}, {del Peloso}, {del Pozo}, {Delbo},
  {Delgado}, {Delgado}, {Delisle}, {Di Matteo}, {Diakite}, {Diener},
  {Distefano}, {Dolding}, {Eappachen}, {Edvardsson}, {Enke}, {Esquej}, {Fabre},
  {Fabrizio}, {Faigler}, {Fedorets}, {Fernique}, {Fienga}, {Figueras},
  {Fouron}, {Fragkoudi}, {Fraile}, {Franke}, {Gai}, {Garabato},
  {Garcia-Gutierrez}, {Garc{\'\i}a-Torres}, {Garofalo}, {Gavras}, {Gerlach},
  {Geyer}, {Giacobbe}, {Gilmore}, {Girona}, {Giuffrida}, {Gomel}, {Gomez},
  {Gonzalez-Santamaria}, {Gonz{\'a}lez-Vidal}, {Granvik},
  {Guti{\'e}rrez-S{\'a}nchez}, {Guy}, {Hauser}, {Haywood}, {Helmi}, {Hidalgo},
  {Hilger}, {H{\l}adczuk}, {Hobbs}, {Holland}, {Huckle}, {Jasniewicz},
  {Jonker}, {Juaristi Campillo}, {Julbe}, {Karbevska}, {Kervella}, {Khanna},
  {Kochoska}, {Kontizas}, {Kordopatis}, {Korn}, {Kostrzewa-Rutkowska},
  {Kruszy{\'n}ska}, {Lambert}, {Lanza}, {Lasne}, {Le Campion}, {Le Fustec},
  {Lebreton}, {Lebzelter}, {Leccia}, {Leclerc}, {Lecoeur-Taibi}, {Liao},
  {Licata}, {Lindstr{\o}m}, {Lister}, {Livanou}, {Lobel}, {Madrero Pardo},
  {Managau}, {Mann}, {Marchant}, {Marconi}, {Marcos Santos}, {Marinoni},
  {Marocco}, {Marshall}, {Martin Polo}, {Mart{\'\i}n-Fleitas}, {Masip},
  {Massari}, {Mastrobuono-Battisti}, {Mazeh}, {McMillan}, {Messina},
  {Michalik}, {Millar}, {Mints}, {Molina}, {Molinaro}, {Moln{\'a}r},
  {Montegriffo}, {Mor}, {Morbidelli}, {Morel}, {Morris}, {Mulone}, {Munoz},
  {Muraveva}, {Murphy}, {Musella}, {Noval}, {Ord{\'e}novic}, {Orr{\`u}},
  {Osinde}, {Pagani}, {Pagano}, {Palaversa}, {Palicio}, {Panahi}, {Pawlak},
  {Pe{\~n}alosa Esteller}, {Penttil{\"a}}, {Piersimoni}, {Pineau}, {Plachy},
  {Plum}, {Poggio}, {Poretti}, {Poujoulet}, {Pr{\v{s}}a}, {Pulone}, {Racero},
  {Ragaini}, {Rainer}, {Raiteri}, {Rambaux}, {Ramos}, {Ramos-Lerate}, {Re
  Fiorentin}, {Regibo}, {Reyl{\'e}}, {Ripepi}, {Riva}, {Rixon}, {Robichon},
  {Robin}, {Roelens}, {Rohrbasser}, {Romero-G{\'o}mez}, {Rowell}, {Royer},
  {Rybicki}, {Sadowski}, {Sagrist{\`a} Sell{\'e}s}, {Sahlmann}, {Salgado},
  {Salguero}, {Samaras}, {Sanchez Gimenez}, {Sanna}, {Santove{\~n}a},
  {Sarasso}, {Schultheis}, {Sciacca}, {Segol}, {Segovia}, {S{\'e}gransan},
  {Semeux}, {Shahaf}, {Siddiqui}, {Siebert}, {Siltala}, {Slezak}, {Smart},
  {Solano}, {Solitro}, {Souami}, {Souchay}, {Spagna}, {Spoto}, {Steele},
  {Steidelm{\"u}ller}, {Stephenson}, {S{\"u}veges}, {Szabados}, {Szegedi-Elek},
  {Taris}, {Tauran}, {Taylor}, {Teixeira}, {Thuillot}, {Tonello}, {Torra},
  {Torra}, {Turon}, {Unger}, {Vaillant}, {van Dillen}, {Vanel}, {Vecchiato},
  {Viala}, {Vicente}, {Voutsinas}, {Weiler}, {Wevers}, {Wyrzykowski}, {Yoldas},
  {Yvard}, {Zhao}, {Zorec}, {Zucker}, {Zurbach}, \&
  {Zwitter}}]{2021A&A...649A...1G}
{Gaia Collaboration}, {Brown}, A.~G.~A., {Vallenari}, A., {et~al.} 2021, \aap,
  649, A1, \dodoi{10.1051/0004-6361/202039657}

\bibitem[{{Gordon} {et~al.}(2023{\natexlab{a}}){Gordon}, {Clayton}, {Decleir},
  {Fitzpatrick}, {Massa}, {Misselt}, \& {Tollerud}}]{2023arXiv230401991G}
{Gordon}, K.~D., {Clayton}, G.~C., {Decleir}, M., {et~al.} 2023{\natexlab{a}},
  arXiv e-prints, arXiv:2304.01991, \dodoi{10.48550/arXiv.2304.01991}

\bibitem[{{Gordon} {et~al.}(2023{\natexlab{b}}){Gordon}, {Clayton}, {Decleir},
  {Fitzpatrick}, {Massa}, {Misselt}, \& {Tollerud}}]{2023ApJ...950...86G}
---. 2023{\natexlab{b}}, \apj, 950, 86, \dodoi{10.3847/1538-4357/accb59}

\bibitem[{{Irfan} {et~al.}(2019){Irfan}, {Bobin}, {Miville-Desch{\^e}nes}, \&
  {Grenier}}]{2019A&A...623A..21I}
{Irfan}, M.~O., {Bobin}, J., {Miville-Desch{\^e}nes}, M.-A., \& {Grenier}, I.
  2019, \aap, 623, A21, \dodoi{10.1051/0004-6361/201834394}

\bibitem[{{K{\"o}hler} {et~al.}(2012){K{\"o}hler}, {Stepnik}, {Jones},
  {Guillet}, {Abergel}, {Ristorcelli}, \& {Bernard}}]{2012A&A...548A..61K}
{K{\"o}hler}, M., {Stepnik}, B., {Jones}, A.~P., {et~al.} 2012, \aap, 548, A61,
  \dodoi{10.1051/0004-6361/201218975}

\bibitem[{{Liu} {et~al.}(2014){Liu}, {Yuan}, {Huo}, {Deng}, {Hou}, {Zhao},
  {Zhao}, {Shi}, {Luo}, {Xiang}, {Zhang}, {Huang}, \&
  {Zhang}}]{2014IAUS..298..310L}
{Liu}, X.~W., {Yuan}, H.~B., {Huo}, Z.~Y., {et~al.} 2014, in Setting the scene
  for Gaia and LAMOST, ed. S.~{Feltzing}, G.~{Zhao}, N.~A. {Walton}, \&
  P.~{Whitelock}, Vol. 298, 310--321, \dodoi{10.1017/S1743921313006510}

\bibitem[{{Luo} {et~al.}(2015){Luo}, {Zhao}, {Zhao}, {Deng}, {Liu}, {Jing},
  {Wang}, {Zhang}, {Shi}, {Cui}, {Chu}, {Li}, {Bai}, {Wu}, {Cai}, {Cao}, {Cao},
  {Carlin}, {Chen}, {Chen}, {Chen}, {Chen}, {Chen}, {Chen}, {Chen},
  {Christlieb}, {Chu}, {Cui}, {Dong}, {Du}, {Fan}, {Feng}, {Fu}, {Gao}, {Gong},
  {Gu}, {Guo}, {Han}, {He}, {Hou}, {Hou}, {Hou}, {Hu}, {Hu}, {Hu}, {Huo},
  {Jia}, {Jiang}, {Jiang}, {Jiang}, {Jin}, {Kong}, {Kong}, {Lei}, {Li}, {Li},
  {Li}, {Li}, {Li}, {Li}, {Li}, {Li}, {Li}, {Li}, {Li}, {Li}, {Liang}, {Lin},
  {Liu}, {Liu}, {Liu}, {Liu}, {Lu}, {Luo}, {Mao}, {Newberg}, {Ni}, {Qi}, {Qi},
  {Shen}, {Shi}, {Song}, {Song}, {Su}, {Su}, {Tang}, {Tao}, {Tian}, {Wang},
  {Wang}, {Wang}, {Wang}, {Wang}, {Wang}, {Wang}, {Wang}, {Wang}, {Wang},
  {Wang}, {Wang}, {Wang}, {Wang}, {Wang}, {Wang}, {Wang}, {Wang}, {Wang},
  {Wang}, {Wei}, {Wei}, {Wu}, {Wu}, {Wu}, {Wu}, {Xing}, {Xu}, {Xu}, {Xu},
  {Yan}, {Yang}, {Yang}, {Yang}, {Yang}, {Yao}, {Yu}, {Yuan}, {Yuan}, {Yuan},
  {Yuan}, {Zhai}, {Zhang}, {Zhang}, {Zhang}, {Zhang}, {Zhang}, {Zhang},
  {Zhang}, {Zhang}, {Zhao}, {Zhou}, {Zhou}, {Zhu}, {Zhu}, {Zou}, \&
  {Zuo}}]{2015RAA....15.1095L}
{Luo}, A.~L., {Zhao}, Y.-H., {Zhao}, G., {et~al.} 2015, Research in Astronomy
  and Astrophysics, 15, 1095, \dodoi{10.1088/1674-4527/15/8/002}

\bibitem[{{Ma{\'\i}z Apell{\'a}niz} {et~al.}(2014){Ma{\'\i}z Apell{\'a}niz},
  {Evans}, {Barb{\'a}}, {Gr{\"a}fener}, {Bestenlehner}, {Crowther},
  {Garc{\'\i}a}, {Herrero}, {Sana}, {Sim{\'o}n-D{\'\i}az}, {Taylor}, {van
  Loon}, {Vink}, \& {Walborn}}]{2014A&A...564A..63M}
{Ma{\'\i}z Apell{\'a}niz}, J., {Evans}, C.~J., {Barb{\'a}}, R.~H., {et~al.}
  2014, \aap, 564, A63, \dodoi{10.1051/0004-6361/201423439}

\bibitem[{{Martin} {et~al.}(2005){Martin}, {Fanson}, {Schiminovich},
  {Morrissey}, {Friedman}, {Barlow}, {Conrow}, {Grange}, {Jelinsky},
  {Milliard}, {Siegmund}, {Bianchi}, {Byun}, {Donas}, {Forster}, {Heckman},
  {Lee}, {Madore}, {Malina}, {Neff}, {Rich}, {Small}, {Surber}, {Szalay},
  {Welsh}, \& {Wyder}}]{2005ApJ...619L...1M}
{Martin}, D.~C., {Fanson}, J., {Schiminovich}, D., {et~al.} 2005, \apjl, 619,
  L1, \dodoi{10.1086/426387}

\bibitem[{{O'Donnell}(1994)}]{1994ApJ...422..158O}
{O'Donnell}, J.~E. 1994, \apj, 422, 158, \dodoi{10.1086/173713}

\bibitem[{{Planck Collaboration} {et~al.}(2016){Planck Collaboration}, {Adam},
  {Ade}, {Aghanim}, {Alves}, {Arnaud}, {Ashdown}, {Aumont}, {Baccigalupi},
  {Banday}, {Barreiro}, {Bartlett}, {Bartolo}, {Battaner}, {Benabed},
  {Beno{\^\i}t}, {Benoit-L{\'e}vy}, {Bernard}, {Bersanelli}, {Bielewicz},
  {Bock}, {Bonaldi}, {Bonavera}, {Bond}, {Borrill}, {Bouchet}, {Boulanger},
  {Bucher}, {Burigana}, {Butler}, {Calabrese}, {Cardoso}, {Catalano},
  {Challinor}, {Chamballu}, {Chary}, {Chiang}, {Christensen}, {Clements},
  {Colombi}, {Colombo}, {Combet}, {Couchot}, {Coulais}, {Crill}, {Curto},
  {Cuttaia}, {Danese}, {Davies}, {Davis}, {de Bernardis}, {de Rosa}, {de
  Zotti}, {Delabrouille}, {D{\'e}sert}, {Dickinson}, {Diego}, {Dole},
  {Donzelli}, {Dor{\'e}}, {Douspis}, {Ducout}, {Dupac}, {Efstathiou}, {Elsner},
  {En{\ss}lin}, {Eriksen}, {Falgarone}, {Fergusson}, {Finelli}, {Forni},
  {Frailis}, {Fraisse}, {Franceschi}, {Frejsel}, {Galeotta}, {Galli}, {Ganga},
  {Ghosh}, {Giard}, {Giraud-H{\'e}raud}, {Gjerl{\o}w}, {Gonz{\'a}lez-Nuevo},
  {G{\'o}rski}, {Gratton}, {Gregorio}, {Gruppuso}, {Gudmundsson}, {Hansen},
  {Hanson}, {Harrison}, {Helou}, {Henrot-Versill{\'e}},
  {Hern{\'a}ndez-Monteagudo}, {Herranz}, {Hildebrandt}, {Hivon}, {Hobson},
  {Holmes}, {Hornstrup}, {Hovest}, {Huffenberger}, {Hurier}, {Jaffe}, {Jaffe},
  {Jones}, {Juvela}, {Keih{\"a}nen}, {Keskitalo}, {Kisner}, {Kneissl},
  {Knoche}, {Kunz}, {Kurki-Suonio}, {Lagache}, {L{\"a}hteenm{\"a}ki},
  {Lamarre}, {Lasenby}, {Lattanzi}, {Lawrence}, {Le Jeune}, {Leahy},
  {Leonardi}, {Lesgourgues}, {Levrier}, {Liguori}, {Lilje}, {Linden-V{\o}rnle},
  {L{\'o}pez-Caniego}, {Lubin}, {Mac{\'\i}as-P{\'e}rez}, {Maggio}, {Maino},
  {Mandolesi}, {Mangilli}, {Maris}, {Marshall}, {Martin},
  {Mart{\'\i}nez-Gonz{\'a}lez}, {Masi}, {Matarrese}, {McGehee}, {Meinhold},
  {Melchiorri}, {Mendes}, {Mennella}, {Migliaccio}, {Mitra},
  {Miville-Desch{\^e}nes}, {Moneti}, {Montier}, {Morgante}, {Mortlock}, {Moss},
  {Munshi}, {Murphy}, {Naselsky}, {Nati}, {Natoli}, {Netterfield},
  {N{\o}rgaard-Nielsen}, {Noviello}, {Novikov}, {Novikov}, {Orlando},
  {Oxborrow}, {Paci}, {Pagano}, {Pajot}, {Paladini}, {Paoletti}, {Partridge},
  {Pasian}, {Patanchon}, {Pearson}, {Perdereau}, {Perotto}, {Perrotta},
  {Pettorino}, {Piacentini}, {Piat}, {Pierpaoli}, {Pietrobon}, {Plaszczynski},
  {Pointecouteau}, {Polenta}, {Pratt}, {Pr{\'e}zeau}, {Prunet}, {Puget},
  {Rachen}, {Reach}, {Rebolo}, {Reinecke}, {Remazeilles}, {Renault}, {Renzi},
  {Ristorcelli}, {Rocha}, {Rosset}, {Rossetti}, {Roudier},
  {Rubi{\~n}o-Mart{\'\i}n}, {Rusholme}, {Sandri}, {Santos}, {Savelainen},
  {Savini}, {Scott}, {Seiffert}, {Shellard}, {Spencer}, {Stolyarov}, {Stompor},
  {Strong}, {Sudiwala}, {Sunyaev}, {Sutton}, {Suur-Uski}, {Sygnet}, {Tauber},
  {Terenzi}, {Toffolatti}, {Tomasi}, {Tristram}, {Tucci}, {Tuovinen}, {Umana},
  {Valenziano}, {Valiviita}, {Van Tent}, {Vielva}, {Villa}, {Wade}, {Wandelt},
  {Wehus}, {Wilkinson}, {Yvon}, {Zacchei}, \& {Zonca}}]{2016A&A...594A..10P}
{Planck Collaboration}, {Adam}, R., {Ade}, P.~A.~R., {et~al.} 2016, \aap, 594,
  A10, \dodoi{10.1051/0004-6361/201525967}

\bibitem[{{Rodrigo} \& {Solano}(2020)}]{2020sea..confE.182R}
{Rodrigo}, C., \& {Solano}, E. 2020, in XIV.0 Scientific Meeting (virtual) of
  the Spanish Astronomical Society, 182

\bibitem[{{Savage} \& {Mathis}(1979)}]{1979ARA&A..17...73S}
{Savage}, B.~D., \& {Mathis}, J.~S. 1979, \araa, 17, 73,
  \dodoi{10.1146/annurev.aa.17.090179.000445}

\bibitem[{{Schlafly} \& {Finkbeiner}(2011)}]{2011ApJ...737..103S}
{Schlafly}, E.~F., \& {Finkbeiner}, D.~P. 2011, \apj, 737, 103,
  \dodoi{10.1088/0004-637X/737/2/103}

\bibitem[{{Schlafly} {et~al.}(2010){Schlafly}, {Finkbeiner}, {Schlegel},
  {Juri{\'c}}, {Ivezi{\'c}}, {Gibson}, {Knapp}, \&
  {Weaver}}]{2010ApJ...725.1175S}
{Schlafly}, E.~F., {Finkbeiner}, D.~P., {Schlegel}, D.~J., {et~al.} 2010, \apj,
  725, 1175, \dodoi{10.1088/0004-637X/725/1/1175}

\bibitem[{{Schlafly} {et~al.}(2016){Schlafly}, {Meisner}, {Stutz},
  {Kainulainen}, {Peek}, {Tchernyshyov}, {Rix}, {Finkbeiner}, {Covey}, {Green},
  {Bell}, {Burgett}, {Chambers}, {Draper}, {Flewelling}, {Hodapp}, {Kaiser},
  {Magnier}, {Martin}, {Metcalfe}, {Wainscoat}, \&
  {Waters}}]{2016ApJ...821...78S}
{Schlafly}, E.~F., {Meisner}, A.~M., {Stutz}, A.~M., {et~al.} 2016, \apj, 821,
  78, \dodoi{10.3847/0004-637X/821/2/78}

\bibitem[{{Schlegel} {et~al.}(1998){Schlegel}, {Finkbeiner}, \&
  {Davis}}]{1998ApJ...500..525S}
{Schlegel}, D.~J., {Finkbeiner}, D.~P., \& {Davis}, M. 1998, \apj, 500, 525,
  \dodoi{10.1086/305772}

\bibitem[{{Skrutskie} {et~al.}(2006){Skrutskie}, {Cutri}, {Stiening},
  {Weinberg}, {Schneider}, {Carpenter}, {Beichman}, {Capps}, {Chester},
  {Elias}, {Huchra}, {Liebert}, {Lonsdale}, {Monet}, {Price}, {Seitzer},
  {Jarrett}, {Kirkpatrick}, {Gizis}, {Howard}, {Evans}, {Fowler}, {Fullmer},
  {Hurt}, {Light}, {Kopan}, {Marsh}, {McCallon}, {Tam}, {Van Dyk}, \&
  {Wheelock}}]{2006AJ....131.1163S}
{Skrutskie}, M.~F., {Cutri}, R.~M., {Stiening}, R., {et~al.} 2006, \aj, 131,
  1163, \dodoi{10.1086/498708}

\bibitem[{{Snow} \& {McCall}(2006)}]{2006ARA&A..44..367S}
{Snow}, T.~P., \& {McCall}, B.~J. 2006, \araa, 44, 367,
  \dodoi{10.1146/annurev.astro.43.072103.150624}

\bibitem[{{Sun} {et~al.}(2022){Sun}, {Yuan}, \& {Chen}}]{2022ApJS..260...17S}
{Sun}, Y., {Yuan}, H., \& {Chen}, B. 2022, \apjs, 260, 17,
  \dodoi{10.3847/1538-4365/ac642f}

\bibitem[{Virtanen {et~al.}(2020)Virtanen, Gommers, Oliphant, Haberland, Reddy,
  Cournapeau, Burovski, Peterson, Weckesser, Bright, {van der Walt}, Brett,
  Wilson, Millman, Mayorov, Nelson, Jones, Kern, Larson, Carey, Polat, Feng,
  Moore, {VanderPlas}, Laxalde, Perktold, Cimrman, Henriksen, Quintero, Harris,
  Archibald, Ribeiro, Pedregosa, {van Mulbregt}, \& {SciPy 1.0
  Contributors}}]{2020SciPy-NMeth}
Virtanen, P., Gommers, R., Oliphant, T.~E., {et~al.} 2020, Nature Methods, 17,
  261, \dodoi{10.1038/s41592-019-0686-2}

\bibitem[{{Vrba} \& {Rydgren}(1984)}]{1984ApJ...283..123V}
{Vrba}, F.~J., \& {Rydgren}, A.~E. 1984, \apj, 283, 123, \dodoi{10.1086/162281}

\bibitem[{{Wang} {et~al.}(2017){Wang}, {Jiang}, {Zhao}, {Chen}, \& {de
  Grijs}}]{2017ApJ...848..106W}
{Wang}, S., {Jiang}, B.~W., {Zhao}, H., {Chen}, X., \& {de Grijs}, R. 2017,
  \apj, 848, 106, \dodoi{10.3847/1538-4357/aa8db7}

\bibitem[{{Weingartner} \& {Draine}(2001)}]{2001ApJ...548..296W}
{Weingartner}, J.~C., \& {Draine}, B.~T. 2001, \apj, 548, 296,
  \dodoi{10.1086/318651}

\bibitem[{{Welty} \& {Fowler}(1992)}]{1992ApJ...393..193W}
{Welty}, D.~E., \& {Fowler}, J.~R. 1992, \apj, 393, 193, \dodoi{10.1086/171497}

\bibitem[{{Wright} {et~al.}(2010){Wright}, {Eisenhardt}, {Mainzer}, {Ressler},
  {Cutri}, {Jarrett}, {Kirkpatrick}, {Padgett}, {McMillan}, {Skrutskie},
  {Stanford}, {Cohen}, {Walker}, {Mather}, {Leisawitz}, {Gautier}, {McLean},
  {Benford}, {Lonsdale}, {Blain}, {Mendez}, {Irace}, {Duval}, {Liu}, {Royer},
  {Heinrichsen}, {Howard}, {Shannon}, {Kendall}, {Walsh}, {Larsen}, {Cardon},
  {Schick}, {Schwalm}, {Abid}, {Fabinsky}, {Naes}, \&
  {Tsai}}]{2010AJ....140.1868W}
{Wright}, E.~L., {Eisenhardt}, P. R.~M., {Mainzer}, A.~K., {et~al.} 2010, \aj,
  140, 1868, \dodoi{10.1088/0004-6256/140/6/1868}

\bibitem[{{Wu} {et~al.}(2011){Wu}, {Luo}, {Li}, {Shi}, {Prugniel}, {Liang},
  {Zhao}, {Zhang}, {Bai}, {Wei}, {Dong}, {Zhang}, \&
  {Chen}}]{2011RAA....11..924W}
{Wu}, Y., {Luo}, A.~L., {Li}, H.-N., {et~al.} 2011, Research in Astronomy and
  Astrophysics, 11, 924, \dodoi{10.1088/1674-4527/11/8/006}

\bibitem[{{Xiang} {et~al.}(2022){Xiang}, {Rix}, {Ting}, {Kudritzki}, {Conroy},
  {Zari}, {Shi}, {Przybilla}, {Ramirez-Tannus}, {Tkachenko}, {Gebruers}, \&
  {Liu}}]{2022A&A...662A..66X}
{Xiang}, M., {Rix}, H.-W., {Ting}, Y.-S., {et~al.} 2022, \aap, 662, A66,
  \dodoi{10.1051/0004-6361/202141570}

\bibitem[{{Yuan} {et~al.}(2013){Yuan}, {Liu}, \& {Xiang}}]{2013MNRAS.430.2188Y}
{Yuan}, H.~B., {Liu}, X.~W., \& {Xiang}, M.~S. 2013, \mnras, 430, 2188,
  \dodoi{10.1093/mnras/stt039}

\bibitem[{{Zhang} \& {Yuan}(2023)}]{2023ApJS..264...14Z}
{Zhang}, R., \& {Yuan}, H. 2023, \apjs, 264, 14,
  \dodoi{10.3847/1538-4365/ac9dfa}

\bibitem[{{Zhao} {et~al.}(2012){Zhao}, {Zhao}, {Chu}, {Jing}, \&
  {Deng}}]{2012RAA....12..723Z}
{Zhao}, G., {Zhao}, Y.-H., {Chu}, Y.-Q., {Jing}, Y.-P., \& {Deng}, L.-C. 2012,
  Research in Astronomy and Astrophysics, 12, 723,
  \dodoi{10.1088/1674-4527/12/7/002}

\end{thebibliography}
\bibliographystyle{aasjournal}



\end{document}